\documentclass[journal]{IEEEtran}

\usepackage[cmex10,intlimits]{amsmath}
\usepackage{amsfonts}
\usepackage{color,colortbl}
\usepackage[pdftex]{graphicx}
\usepackage{tikz}
\usepackage{cite}
\usepackage{url}
\usepackage{hyperref}
\usepackage{booktabs} 
\usepackage{multirow}
\usepackage{cases} 
\usepackage[nolist]{acronym}
\usepackage[normalem]{ulem}
\usepackage{subfigure}
\usepackage{units}
\usepackage{etoolbox}
\usepackage{animate}

\newtoggle{ARXIV}
\toggletrue{ARXIV}	

\usepackage{mathtools}
\DeclarePairedDelimiter\ceil{\lceil}{\rceil}

\hypersetup{
	colorlinks=true, 
	linkcolor=blue!70!black, 
	citecolor=red!70!black, 
	filecolor=blue!70!black, 
	urlcolor=blue!70!black, 
}

\newcommand{\J}{\mathrm{j}}                 

\providecommand{\D}{\,\mathrm{d}}           
\providecommand{\V}[1]{\boldsymbol{#1}}     
\providecommand{\M}[1]{\mathbf{#1}}         
\providecommand{\Idmat}{\M{1}}              
\newcommand{\UV}[1]{\hat{\V{#1}}}           
\newcommand{\OP}[1]{{\mathcal{#1}}}         

\providecommand{\herm}{\mathrm{H}} 
\providecommand{\trans}{\mathrm{T}}

\providecommand{\basfcn}{\V{\psi}}       
\providecommand{\srcRegion}{\Omega} 
\providecommand{\Ivec}{\M{I}}

\providecommand{\ZVAC}{Z_0}  
\providecommand{\Prad}{P_\mathrm{r}}

\providecommand{\UFCN}[3]{\M{u}_{#1}^{\left( #2 \right)} \left(#3\right)}
\providecommand{\RFCN}[3]{\mathrm{R}_{#1}^{\left( #2 \right)} \left(#3\right)}
\providecommand{\YFCN}[2]{\M{Y}_{#1}\left(#2\right)}

\providecommand{\YFCNs}[2]{\mathrm{Y}_{#1}\left(#2\right)}

\providecommand{\PFCNn}[3]{\tilde{\mathrm{P}}_{#1}^{#2}\left(#3\right)}
\providecommand{\ConstB}{b_l}

\providecommand{\Lmax}{L}
\providecommand{\Nquad}{N_q}
\providecommand{\Npsi}{N_{\psi}}
\providecommand{\Nu}{N_{\alpha}}
\providecommand{\Smat}{\M{S}}

\newcommand\figwidth{8.8} 
\graphicspath{{figures/}}
\newcommand{\ie}{\textit{i.e.}{}}
\newcommand{\eg}{\textit{e.g.}{}}
\newcommand{\cf}{\textit{cf.}{}}

\iftoggle{ARXIV}{%
\newcommand{\TRa}[1]{#1}   
\newcommand{\TRsa}[1]{}   
}
{
\newcommand{\TRa}[1]{#1}   
\newcommand{\TRsa}[1]{}   
}

\newacro{MoM}[MoM]{method of moments}
\newacro{CM}[CM]{characteristic mode}
\newacro{PEC}[PEC]{perfect electric conductor}
\newacro{EP}[EP]{eigenvalue problem}
\newacro{GEP}[GEP]{generalized eigenvalue problem}
\newacro{EFIE}[EFIE]{electric field integral equation}
\newacro{SVD}[SVD]{singular value decomposition}
\newacro{dof}[d-o-f]{\mbox{degrees-of-freedom}}

\begin{document}

\title{Accurate and Efficient Evaluation of Characteristic~Modes}
\author{Doruk~Tayli,~\IEEEmembership{Student~Member,~IEEE,}
	Miloslav~Capek,~\IEEEmembership{Senior~Member,~IEEE,}
	Lamyae~Akrou,	
	Vit~Losenicky,	
	Lukas~Jelinek,
	and~Mats~Gustafsson,~\IEEEmembership{Senior~Member,~IEEE}
\thanks{Manuscript received  \today; revised \today.
This work was supported by the Swedish Foundation for Strategic Research (SSF) under the program Applied Mathematics and the project Complex analysis and convex optimization for EM design, and by the Czech Science Foundation under project No. 15-10280Y.}
\thanks{D.~Tayli and M.~Gustafsson are with the Department of Electrical and Information Technology,
	Lund University, 221~00 Lund, Sweden (e-mail: \{doruk.tayli,mats.gustafsson\}@eit.lth.se).}
\thanks{M.~Capek, V.~Losenicky and L.~Jelinek are with the Department of Electromagnetic Field, Faculty of Electrical Engineering, Czech Technical University in Prague, Technicka 2, 166~27 Prague, Czech Republic (e-mail: \{miloslav.capek,losenvit,lukas.jelinek\}@fel.cvut.cz).}
\thanks{L.~Akrou is with the Department of Electrical and Computer Engineering, Faculty of Sciences and Technology, University of Coimbra, Polo II, Pinhal de Marrocos, 3030-290 Coimbra, Portugal (e-mail: lakrou@co.it.pt).}
}

\markboth{Journal of \LaTeX\ Class Files,~Vol.~XX, No.~XX, \today}
{Tayli et al.: Accurate and Efficient Evaluation of Characteristic Modes}

\maketitle

\begin{abstract}
A new method to improve the accuracy and efficiency of \acf{CM} decomposition for perfectly conducting bodies is presented.
The method uses the expansion of the Green dyadic in spherical vector waves. This expansion is utilized in the \acf{MoM} solution of the \ac{EFIE} to factorize the real part of the impedance matrix. The factorization is then employed in the computation of \acp{CM}, which improves the accuracy as well as the computational speed. An additional benefit is a rapid computation of far fields. The method can easily be integrated into existing \ac{MoM} solvers. Several structures are investigated illustrating the improved accuracy and performance of the new method.
\end{abstract}


\IEEEpeerreviewmaketitle

\section{Introduction}
\label{sec:Intro}

\IEEEPARstart{T}{he} \acf{MoM} solution to electromagnetic field integral equations was introduced by Harrington~\cite{Harrington_FieldComputationByMoM} and has prevailed as a standard in solving open (radiating) electromagnetic problems~\cite{Sadiku_NumericalTechniques}. While memory-demanding, \ac{MoM} represents operators as matrices (notably the impedance matrix \cite{Harrington_FieldComputationByMoM}) allowing for direct inversion and modal decompositions~\cite{GolubVanLoan_MatrixComputations}. The latter option is becoming increasingly popular, mainly due to \acf{CM} decomposition \cite{Harrington+Mautz1971}, a leading formalism in antenna shape and feeding synthesis ~\cite{YangAdams_SystematicShapeOptimizationOfSymmetricMIMOAntennasUsingCM, CapekHazdraEichler_AMethodForTheEvaluationOfRadiationQBasedOnModalApproach}, determination of optimal currents ~\cite{CapekJelinek_OptimalCompositionOfModalCurrentsQ, GustafssonTayliEhrenborgEtAl_AntennaCurrentOptimizationUsingMatlabAndCVX}, and performance evaluation~\cite{VogelEtAl_CManalysis_PuttingPhysicsBackIntoSimulation}. 

Utilization of \ac{CM} decomposition is especially efficient when dealing with electrically small antennas~\cite{Chen_2014_UAV_TCM}, particularly if they are made solely of \ac{PEC}, for which only a small number of modes are needed to describe their radiation behavior. Yet, the real part of the impedance matrix is indefinite as it is computed with finite precision~\mbox{\cite{HarringtonMautz_ComputationOfCharacteristicModesForConductingBodies,CapekEtAl_ValidatingCMsolvers}}.
The aforementioned deficiency is resolved in this paper by a two-step procedure. First, the real part of the impedance matrix is constructed using spherical wave expansion of the dyadic Green function~\cite{Kristensson2016}. This makes it possible to decompose the real part of the impedance matrix as a product of a spherical modes projection matrix with its hermitian conjugate. The second step consists of reformulating the modal decomposition so that only the standalone spherical modes projection matrix is involved preserving the numerical dynamics\footnote{The numerical dynamic is defined as the largest characteristic eigenvalue.}. 

The proposed method significantly accelerates the computation of \acp{CM} as well as of the real part of the impedance matrix. Moreover, it is possible to recover \acp{CM} using lower precision floating point arithmetic, which reduces memory use and speeds up arithmetic operations if hardware vectorization is exploited~\cite{GolubVanLoan_MatrixComputations}. An added benefit is the efficient computation of far field patterns using spherical vector harmonics.

The projection on spherical waves in the proposed method introduces several appealing properties. First is an easy monitoring of the numerical dynamics of the matrix, since the different spherical waves occupy separate rows in the projection matrix. Second is the possibility to compute a positive semidefinite impedance matrix which plays important role in an optimal design~\mbox{\cite{Gustafsson_OptimalAntennaCurrentsForQsuperdirectivityAndRP,GustafssonTayliEhrenborgEtAl_AntennaCurrentOptimizationUsingMatlabAndCVX}}. A final benefit is the superposition of modes.~\mbox{\cite{CapekHazdraEichler_AMethodForTheEvaluationOfRadiationQBasedOnModalApproach}}.

The paper is organized as follows. The construction of the impedance matrix using classical procedure is briefly reviewed in Section~\ref{sec:EFIE} and the proposed procedure is presented in Section~\ref{sec:SPH}. Numerical aspects of evaluating the impedance matrix are discussed in Section~\ref{sec:Zaspects}. In Section~\ref{sec:Decomposition}, the spherical modes projection matrix is utilized to reformulate modal decomposition techniques, namely the evaluation of radiation modes in Section~\ref{sec:RadModes} and \acp{CM} in Section~\ref{sec:CMmodes}. These two applications cover both the standard and generalized eigenvalue problems. The advantages of the proposed procedure are demonstrated on a series of practical examples in this section. Various aspects of the proposed method are discussed in Section~\ref{sec:Disc} and the paper is concluded in Section~\ref{sec:Concl}.

\section{Evaluation of Impedance Matrix}
\label{sec:Method}

This paper investigates mode decompositions for \ac{PEC} structures in free space. The time-harmonic quantities under the convention \mbox{$\boldsymbol{\mathcal{J}} \left(\V{r},t\right) = \mathrm{Re}\left\{\V{J}\left(\V{r},\omega\right) \mathrm{exp}\left(\J \omega t \right)\right\}$}, with $\omega$ being the angular frequency, are used throughout the paper.

\subsection{Method of Moments Implementation of the \acs{EFIE}}
\label{sec:EFIE}

Let us consider the \acf{EFIE}~\cite{Harrington_FieldComputationByMoM} for \ac{PEC} bodies, defined as 
\begin{equation}
\label{eq:ZOperDef}
\mathcal{Z}\left(\V{J}\right)=\OP{R}\left(\V{J}\right)+\J\OP{X}\left(\V{J}\right) = \UV{n} \times \left( \UV{n} \times \V{E} \right),
\end{equation}
with $\mathcal{Z}\left(\V{J}\right)$ being the impedance operator, $\V{E}$ the incident electric field \cite{Harrington_TimeHarmonicElmagField}, $\V{J}$ the current density, $\J$ the imaginary unit, and $\V{\UV{n}}$ the unit normal vector to the PEC surface. The \ac{EFIE}~\eqref{eq:ZOperDef} is explicitly written as
\begin{equation}
\label{eq:Efield}
\UV{n}\times\V{E}\left(\V{r}_2\right)=\J k \ZVAC \UV{n}\times \int_{\srcRegion}  \M{G} \left(\V{r}_1, \V{r}_2\right)\cdot \V{J}  \left(\V{r}_1\right) \D{A}_1,
\end{equation}
where \mbox{$\V{r}_2\in \Omega$}, $k$ is the wave number, $\ZVAC$ the free space impedance, and $\M{G}$ the dyadic Green function for the electric field in free-space defined as~\cite{Kristensson2016,ChewTongHu_IntegralEquationMethodsForElectromagneticAndElasticWaves}
\begin{equation}
\label{eq:DGreenExp}
\M{G} \left(\V{r}_1,\V{r}_2\right) = \left(\Idmat + \frac{1}{k^2}\nabla\nabla\right) \frac{\mathrm{e}^{-\J k \left|\V{r}_1 - \V{r}_2\right|}}{4 \pi \left|\V{r}_1 - \V{r}_2\right|}.
\end{equation}
Here, $\Idmat$ is the identity dyadic, and $\V{r}_1$, $\V{r}_2$ are the source and observation points. The \ac{EFIE}~\eqref{eq:Efield} is solved with the \ac{MoM} by expanding the current density $\V{J} \left(\V{r}\right)$ into real-valued basis functions $\left\{\basfcn_p \left(\V{r}\right)\right\}$ as
\begin{equation}
\label{eq:basFcn}
\V{J} \left(\V{r}\right) \approx \sum_{p=1}^{\Npsi} I_{p} \basfcn_p \left(\V{r}\right)
\end{equation}
and applying Galerkin testing procedure~\cite{ChewTongHu_IntegralEquationMethodsForElectromagneticAndElasticWaves,Gibson2014}. The impedance operator \mbox{$\mathcal{Z}\left(\V{J}\right)$} is expressed as the impedance matrix \mbox{$\M{Z} = \M{R} + \J \M{X} = \left[Z_{pq}\right] \in \mathbb{C}^{\Npsi\times \Npsi}$}, where $\M{R}$ is the resistance matrix, and $\M{X}$ the reactance matrix. The elements of the impedance matrix are
\begin{equation}
\label{eq:Zitem}
Z_{pq} = \J k \ZVAC \int_{\srcRegion} \int_{\srcRegion} \basfcn_p \left(\V{r}_1\right) \cdot \M{G} \left(\V{r}_1,\V{r}_2\right) \cdot \basfcn_q \left(\V{r}_2\right) \D{A}_1 \D{A}_2.
\end{equation}

\subsection{Spherical Wave Expansion of the Green Dyadic}
\label{sec:SPH}
The Green dyadic~\eqref{eq:DGreenExp} that is used to compute the impedance matrix $\M{Z}$ can be expanded in spherical vector waves as
\begin{equation}
\label{eq:DGreenSph}
\M{G} \left(\V{r}_1,\V{r}_2\right) = -\J k \sum\limits_\alpha \UFCN{\alpha}{1}{k\V{r}_<} \UFCN{\alpha}{4}{k\V{r}_>},
\end{equation}
where \mbox{$\V{r}_< = \V{r}_1$} and \mbox{$\V{r}_> = \V{r}_2$} if \mbox{$\lvert\V{r}_1\rvert < \lvert\V{r}_2\rvert$}, and \mbox{$\V{r}_< = \V{r}_2$} and \mbox{$\V{r}_> = \V{r}_1$} if \mbox{$\lvert\V{r}_1\rvert > \lvert\V{r}_2\rvert$}. The regular and outgoing spherical vector waves~\cite{Morse+Feshbach1953b,Waterman1971,Hansen1988,Kristensson2016} are \mbox{$\UFCN{\alpha}{1}{k\V{r}}$} and \mbox{$\UFCN{\alpha}{4}{k\V{r}}$}, see Appendix~\ref{app:SPHwaves}. The mode index $\alpha$ for \mbox{real-valued} vector spherical harmonics is~\cite{Hansen1988,Gustafsson+Nordebo2006c}
\begin{equation}
\alpha\left( \tau, \sigma,  m, l \right) = 2 \left(l^2+l-1+(-1)^{s}m \right) + \tau
\end{equation}
with \mbox{$\tau \in \left\{1,2\right\}$}, \mbox{$m\in\left\{0,\dots,l\right\}$}, \mbox{$l\in\left\{1,\dots,\Lmax\right\}$}, $s=0$ for  even azimuth functions ($\sigma=\mathrm{e}$), and $s=1$ for odd azimuth functions ($\sigma=\mathrm{o}$).
Inserting the expansion of the Green dyadic~\eqref{eq:DGreenSph} into~\eqref{eq:Zitem}, the impedance matrix~$\M{Z}$ becomes
\begin{multline}
\label{eq:Zsph}
Z_{pq} = k^2 \ZVAC \sum_{\alpha}\int_\srcRegion\int_\srcRegion \basfcn_p \left(\V{r}_1\right)\,\cdot\,\UFCN{\alpha}{1}{k\V{r}_<}\\
\UFCN{\alpha}{4}{k\V{r}_>}\,\cdot\,\basfcn_q \left(\V{r}_2\right) \D{A}_1 \D{A}_2.
\end{multline}
For a \ac{PEC} structure the resistive part of~\eqref{eq:Zsph} can be factorized as
\begin{multline}
\label{eq:Ritem}
R_{pq} = k^2 \ZVAC \sum_{\alpha}\int_\srcRegion \basfcn_p \left(\V{r}_1\right)\,\cdot\,\UFCN{\alpha}{1}{k\V{r}_1}\D{A}_1\\
\int_\srcRegion\UFCN{\alpha}{1}{k\V{r}_2}\,\cdot\,\basfcn_q \left(\V{r}_2\right) \D{A}_2,
\end{multline}
where $\UFCN{\alpha}{1}{k\V{r}}=\mathrm{Re}\{\UFCN{\alpha}{4}{k\V{r}}\}$ is used. Reactance matrix,~$\M{X}$, cannot be factorized in a similar way as two separate spherical waves occur.

Resistance matrix can be written in matrix form as
\begin{equation}
\label{eq:Rdecomp}
\M{R} = \Smat^{\trans} \Smat,
\end{equation}
where $^{\trans}$ is the matrix transpose. Individual elements of the matrix $\Smat$ are
\begin{equation}
\label{eq:Sau}
S_{\alpha p} =  k \sqrt{\ZVAC} \int_\srcRegion \basfcn_p \left(\V{r}\right) \, \cdot \, \UFCN{\alpha}{1}{k\V{r}} \D{A}
\end{equation}
and the size of the matrix $\Smat$ is \mbox{$\Nu \times \Npsi$}, where 
\begin{equation}
\label{eq:numberOfSphWaves}
\Nu = 2 \Lmax \left(\Lmax + 2\right)
\end{equation}
is the number of spherical modes and $\Lmax$ the highest order of spherical mode, see Appendix~\ref{app:SPHwaves}. 
For \mbox{complex-valued} vector spherical harmonics~\cite{Hansen1988} the transpose $^\trans$ in~\eqref{eq:Rdecomp} is replaced with the hermitian transpose $^\herm$.
 The individual integrals in~\eqref{eq:Zsph} are in fact related to the \mbox{T-matrix method}~\cite{Waterman1965,Waterman1971}, where the incident and scattered electric fields are expanded using regular and outgoing spherical vector waves, respectively. The factorization~\eqref{eq:DGreenSph} is also used in vector fast multipole algorithm~\cite{Liu+etal2012}.

The radiated \mbox{far-field} $\V{F}\left(\UV{r}\right)$ can conveniently be computed using spherical vector harmonics
\begin{equation}
\label{eq:Ftest3}
\V{F} \left(\UV{r}\right) = \frac{1}{k} \sum_{\alpha} \J^{l-\tau+2} f_{\alpha} \YFCN{\alpha}{\UV{r}},
\end{equation}
where $\YFCN{\alpha}{\UV{r}}$ are the spherical vector harmonics, see Appendix~\ref{app:SPHwaves}. The expansion coefficients $f_{\alpha}$ are given by
\begin{equation}
\label{eq:Ftest4}
\left[ f_{\alpha} \right] = \Smat \Ivec,
\end{equation}
where the column matrix~$\Ivec$ contains the current density coefficients $I_{p}$. 
The total \mbox{time-averaged} radiated power of a lossless antenna can be expressed as a sum of expansion coefficients
\begin{equation}
\label{eq:Prad}
\Prad \approx \frac{1}{2}\Ivec^\herm\M{R}\Ivec = \frac{1}{2}\lvert \Smat\Ivec\rvert^2 = \frac{1}{2}\sum_{\alpha}\lvert f_{\alpha}\rvert^2.
\end{equation}

\subsection{Numerical Considerations}
\label{sec:Zaspects}

The spectrum of the matrices $\M{R}$ and $\M{X}$ differ considerably~\cite{GustafssonTayliEhrenborgEtAl_AntennaCurrentOptimizationUsingMatlabAndCVX,CapekEtAl_ValidatingCMsolvers}. The eigenvalues of the $\M{R}$ matrix decrease exponentially and the number of eigenvalues are corrupted by numerical noise, while this is not the case for the matrix~$\M{X}$. As a result, if the matrix~$\M{R}$ is used in an eigenvalue problem, only a few modes can be extracted. This major limitation can be overcome with the use of the matrix~$\Smat$ in \eqref{eq:Sau}, whose elements vary several order of magnitude, as the result of the increased order of spherical modes with increasing row number. If the matrix~$\M{R}$ is directly computed with the matrix product~\eqref{eq:Rdecomp} or equivalently from matrix produced by~\eqref{eq:Zitem} small values are truncated due to \mbox{floating-point} arithmetic\footnote{As an example to the loss of significance in double precision arithmetic consider the sum \mbox{$1.0+1\times10^{-30}= 1.0$}.}~\cite{Zuras2008,Burden2015}. Subsequently, the spectrum of the matrix~$\M{R}$ should be computed from the matrix~$\Smat$ as presented in Section~\ref{sec:Decomposition}.

\begin{figure}[t]
	\begin{center}
		\includegraphics[width=\figwidth cm]{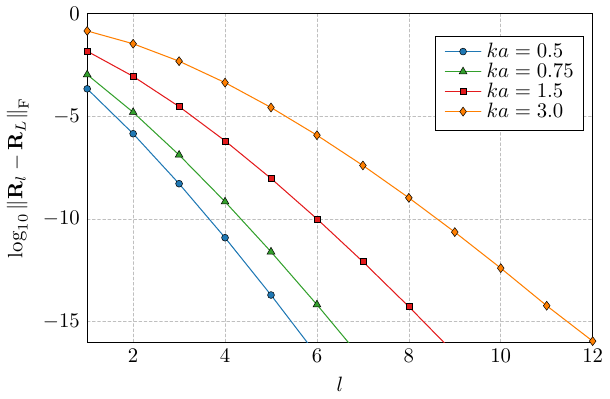}
		\caption{Convergence of the matrix~\mbox{$\M{R}_l = \Smat^{\trans}_l\Smat_l$} to the matrix~\mbox{$\M{R}_L = \Smat^{\trans}_{L}\Smat_{L}$} on the rectangular plate (Example~\hyperref[tab:Examples]{R2}) for different order of spherical modes \mbox{$l=\left\{1,\dots,L\right\}$} and multiple electric sizes \mbox{$ka \in \left\{0.5,0.75,1.5,3.0\right\}$}, with the highest spherical mode order $L = 12$. The superscript $^{\mathrm{F}}$ denotes the Frobenius norm. The convergence is computed with quadruple precision using the mpmath Python library~\cite{mpmath2013}.}
		\label{fig:Plate_Lstudy}
	\end{center}
\end{figure}

The matrix~$\Smat$ also provides a low-rank approximation of the matrix~$\M{R}$, which is the result of the rapid convergence of regular spherical waves. In this paper, the number of used modes in~\eqref{eq:DGreenSph} is truncated using a modified version of the expression in~\cite{Song+Chew2001a}
\begin{equation}
\label{eq:Lmax}
\Lmax=\ceil{ka+7\sqrt[3]{ka}+3},
\end{equation}
where $\Lmax$ is the highest order of spherical mode, $a$ is the radius of the sphere enclosing the scatterer, and $\ceil{.}$ is the ceiling function. The resulting accuracy in all treated cases is satisfactory. The order of spherical modes can be modified to trade between accuracy and computational efficiency, where increasing $\Lmax$ improves the accuracy.
Fig.~\ref{fig:Plate_Lstudy} shows the convergence of the matrix~$\M{R}$ for Example~\hyperref[tab:Examples]{R2}.

Substitution of the spherical vector waves, introduced in Section~\ref{sec:SPH}, separates~\eqref{eq:Zitem} into two separate surface integrals reducing computational complexity. Table~\ref{tab:Speed} presents computation times\footnote{Computations are done on a workstation with i7-3770~CPU @ $3.4$~GHz and $32$~GB~RAM, operating under Windows~7.} of different matrices\footnote{Computation time for the matrix~$\M{X}$ is omitted as it takes longer than the matrix~$\M{R}$, due to Green function singularity.} $\M{Z}$, $\M{R}$, $\Smat$, and $\Smat^{\trans}\Smat$ for the examples given in Table~\ref{tab:Examples}. As expected, the matrix~$\M{Z}$ requires the most computational resources, as it includes both the matrix~$\M{R}$ and~$\M{X}$. The computation of the matrix~$\M{R}$ using \ac{MoM} is faster than the matrix~$\M{Z}$ since the underlying integrals are regular. The computation of the matrix~$\M{R}$ using~\eqref{eq:Rdecomp} takes the least amount of time for most of the examples. The computational gain is notable for structures with more \ac{dof},~$\Npsi$.

\begin{table}[t]
	\begin{center}
		\begin{tabular}{ccccc}
			\toprule
	Example & \multicolumn{4}{c}{Time to assemble matrices in IDA (s)} \\ 
	\cmidrule{2-5}
	(see Table~\ref{tab:Examples}) & $\M{Z}$ & $\M{R}$ & $\Smat$ & $\M{R}=\M{S}^{\trans}\M{S}$ \\ \midrule
			\hyperref[tab:Examples]{S1} & $2.58$ 	& $0.09$ & $0.009$ & $0.011$\\ 
			\hyperref[tab:Examples]{S4} & $14.2$ 	& $1.78$  & $0.039$ & $0.083$\\ 
			\hyperref[tab:Examples]{R3} & $11.1$ 	& $1.11$ & $0.035$ & $0.068$\\ 
			\hyperref[tab:Examples]{H1} & $200$ 	& $54.5$ & $0.236$ & $1.66$\\ \bottomrule
		\end{tabular}
	\end{center}
	\caption{Time to assemble matrices in IDA. Simulation setup for the examples in Table~\ref{tab:Examples}, $\Nquad = 3$ and $\Lmax = 10$ ($\Nu=240$), matrix multiplication $\M{S}^{\trans}\M{S}$ is performed with dgemm from the Intel MKL library~\cite{IntelMKL2017}.}
	\label{tab:Speed}
\end{table}

\section{Modal Decomposition With the Matrix~$\Smat$}
\label{sec:Decomposition}

Modal decomposition using the matrix~$\Smat$ is applied to two structures; a spherical shell of radius~$a$, and a rectangular plate of length~$L$ and width~$W=L/2$
\iftoggle{ARXIV}{%
(App.~\ref{app:Mesh}),
}
{
~\cite{TayliEtAl_AccurateAndEfficientEvaluationofCMs},
}
are presented in Table~\ref{tab:Examples}. Both structures are investigated for different number of \ac{dof}, RWG functions~\cite{RaoWiltonGlisson_ElectromagneticScatteringBySurfacesOfArbitraryShape} are used as the basis functions~$\basfcn_p$. The matrices used in modal decomposition have been computed using in-house solvers AToM~\cite{atom} and IDA~\cite{IDA}, see Appendix~\ref{app:CEMpackages} for details. Results from the commercial electromagnetic solver FEKO~\cite{feko} are also presented for comparison. Computations that require a higher precision than the double precision arithmetic are performed using the mpmath Python library~\cite{mpmath2013}, and the Advanpix Matlab toolbox~\cite{advanpix}. 

\begin{table}[t]
	\begin{center}
		\begin{tabular}{ccccc}
			\toprule
			Structure & Example & $ka$ & $\Npsi$ & $\Nu$ \\ \midrule
			                & S1 & $1/2$ & $750$ & $240$ \\
											& S2 & $1/2$ & $750$ & $880$ \\
			Spherical shell & S3 & $3/2$ & $750$ & $880$ \\
			\iftoggle{ARXIV}{
			Fig.~\ref{fig:SphereNumericalMesh}
			}
			{Fig.~9 in~\cite{TayliEtAl_AccurateAndEfficientEvaluationofCMs}}
											& S4 & $1/2$ & $3330$ & $240$ \\ 
											& S5 & $1/2$ & $3330$ & $880$ \\ \midrule
       	  Rectangular plate & R1 & $1/2$ & $199$ & $510$ \\
			\iftoggle{ARXIV}{%
			Fig.~\ref{fig:RectangleNumericalMesh}
			}
			{Fig.~10 in~\cite{TayliEtAl_AccurateAndEfficientEvaluationofCMs}}
											& R2 & $1/2$ & $655$ & $510$ \\
	    $\left(L/W=2\right)$ & R3 & $1/2$ & $2657$ & $240$ \\ 
											& R4 & $1/2$ & $2657$ & $1920$ \\ \midrule
			Helicopter 			& H1 & $1/2$ & $18898$ & $240$ \\ \iftoggle{ARXIV}{}{} 
			           			& H2 & $7$ & $18898$ & $720$ \\ \bottomrule            
		\end{tabular}
	\end{center}
	\caption{Summary of examples used throughout the paper, $ka$ is the electrical size, $\Npsi$ is the number of basis functions \eqref{eq:basFcn}, and $\Nu$ is number of spherical modes calculated as \eqref{eq:numberOfSphWaves}. The order of the symmetric quadrature rule used to compute the non-singular integrals in \eqref{eq:Zitem} is \mbox{$\Nquad = 3$}~\cite{Dunavant_HighDegreeEfficientGQR}.}
	\label{tab:Examples}
\end{table}
\begin{table}[t]
	\begin{center}
		\begin{tabular}{cccccc}
			\toprule
			& \multicolumn{5}{c}{Number of properly calculated modes} \\ \cmidrule{2-6}
	        Example & \multicolumn{2}{c}{$\M{R}\Ivec = \xi_n \Ivec_n$} & \multicolumn{3}{c}{$\M{X}\Ivec_n = \lambda_n \M{R} \Ivec_n$} \\ \cmidrule{2-6}			
     		(see Table~\ref{tab:Examples}) & \eqref{eq:Rmodes} & \eqref{eq:Smodes} & \eqref{eq:CMred1} & \mbox{$\M{R}=\M{S}^\trans\M{S}$} & \eqref{eq:CMred5} \\ \toprule
	     		\hyperref[tab:Examples]{S2} & $59$ & $284$ & $\mathbf{70}$ $\left(5\right)$ & $96$ $\left(6\right)$ & $\mathbf{284}$ $\left(11\right)$ \\ 
	     		\hyperref[tab:Examples]{S3} & $96$ & $364$ & $\mathbf{105}$ $\left(6\right)$ & $197$ $\left(9\right)$ & $\mathbf{389}$ $\left(13\right)$ \\
	     		\hyperref[tab:Examples]{S5} & $59$ & $311$ & $\mathbf{70}$ $\left(5\right)$ & $96$ $\left(6\right)$ & $\mathbf{306}$ $\left(11\right)$ \\ \midrule
	     		\hyperref[tab:Examples]{R1} & $31$ & $109$ & $\mathbf{29}$ & $35$ & $\mathbf{37}$ \\
	     		\hyperref[tab:Examples]{R2} & $29$ & $117$ & $\mathbf{26}$ & $33$ & $\mathbf{98}$ \\
	     		\hyperref[tab:Examples]{R4} & $28$ & $116$ & $\mathbf{22}$ & $26$ & $\mathbf{98}$ \\ \bottomrule
		\end{tabular}
	\end{center}
	\caption{Comparison of the number of modes correctly found by the classical and the novel methods for examples listed in Table~\ref{tab:Examples}. Columns $2$--$3$ summarize the radiation modes and columns $4$--$6$ summarize the \acp{CM}. Values in parentheses depicts the number of non-degenerated TM and TE modes found on spherical shell. The main outcome of the table, comparison of the \acp{CM} is highlighted by bold type.}
	\label{tab:FoundModes}
\end{table}

\subsection{Radiation Modes}
\label{sec:RadModes}
The eigenvalues for the radiation modes~\cite{SchabBernhard_RadiationAndEnergyStorageCurrentModesOnConductingStructures} are easily found using the eigenvalue problem
\begin{equation}
\label{eq:Rmodes}
\M{R} \Ivec_n = \xi_n \Ivec_n,
\end{equation}
where $\xi_n$ are the eigenvalues of the matrix~$\M{R}$, and $\Ivec_n$ are the eigencurrents. The indefiniteness of the matrix~$\M{R}$ poses a problem in the eigenvalue decomposition~\eqref{eq:Rmodes} as illustrated in~\cite{GustafssonTayliEhrenborgEtAl_AntennaCurrentOptimizationUsingMatlabAndCVX,CapekEtAl_ValidatingCMsolvers}. In this paper we show that the indefiniteness caused by the numerical noise can be bypassed using the matrix $\Smat$. We start with the \ac{SVD} of the matrix $\Smat$
\begin{equation}
\label{eq:SVD}
\M{S} = \M{U} \M{\Lambda} \M{V}^\herm,
\end{equation}
where $\M{U}$ and $\M{V}$ are unitary matrices, and $\M{\Lambda}$ is a diagonal matrix containing singular values of matrix~$\Smat$. Inserting~\eqref{eq:Rdecomp},~\eqref{eq:SVD} into~\eqref{eq:Rmodes} and multiplying from the left with $\M{V}^\herm$ yields
\begin{equation}
\label{eq:Smodes}
\M{\Lambda}^{\herm}\M{\Lambda} \widetilde{\Ivec}_n = \xi_n \widetilde{\Ivec}_n,
\end{equation}
where the eigenvectors are rewritten as \mbox{$\widetilde{\Ivec}_n \equiv \M{V}^\herm\Ivec_n$}, and the eigenvalues are \mbox{$\xi_n=\Lambda_{nn}^2$}. \TRsa{The number of radiation modes is shown in Table~\ref{tab:FoundModes} for all the examples. The proper modes are defined to have $5\%$ deviation from the computation with quadruple precision. For the characteristic eigenvalues of the spherical shell, the correct modes are compared with the analytical values, and a 5\% threshold is selected for the error.} \TRa{A comparison of procedure~\eqref{eq:Rmodes} and~\eqref{eq:Smodes} is shown  in Table~\ref{tab:FoundModes}. For high order~$n$, the classical procedure~\eqref{eq:Rmodes} with double numerical precision yields in unphysical modes with negative eigenvalues~$\xi_n$ (negative radiated power) or with incorrect current profile (as compared to the use of quadruple precision). Using double precision, the number of modes which resemble physical reality (called ``properly calculated modes'' in Table~\ref{tab:FoundModes}) is much higher\footnote{\TRa{Quantitatively, the proper modes in Table~\ref{tab:FoundModes} are defined as those having less than $5 \%$ deviation in eigenvalue~$\xi_n$ as compared to the computation with quadruple precision.}} for the new procedure~\eqref{eq:Smodes}. It is also worth mentioning that the new procedure, by design, always gives positive eigenvalues~$\xi_n$.}



\subsection{Characteristic Modes (CMs)}
\label{sec:CMmodes}

The \ac{GEP} with the matrix~$\M{R}$ on the right hand side, \ie{}, serving as a weighting operator~\cite{Wilkinson_AlgebraicEigenvalueProblem}, is much more involved as the problem cannot be completely substituted by the \ac{SVD}. Yet, the \ac{SVD} of the matrix~$\Smat$ in~\eqref{eq:SVD} plays an important role in \TRsa{the} \ac{CM} decomposition. 

\TRa{The \ac{CM} decomposition is defined as} \TRsa{The \ac{CM} decomposition is defined here with a \ac{GEP} as}
\begin{equation}
\label{eq:CMred1}
\M{X} \Ivec_n = \lambda_n \M{R} \Ivec_n,
\end{equation}
which is known to suffer from the indefiniteness of the matrix~$\M{R}$~\cite{CapekEtAl_ValidatingCMsolvers}, therefore delivering only a limited number of modes. The first step is to represent the solution in a basis of singular vectors $\M{V}$ by substituting the matrix~$\M{R}$ in \eqref{eq:CMred1} as~\eqref{eq:Rdecomp}, with \eqref{eq:SVD} and multiplying~\eqref{eq:CMred1} from the left by the matrix~$\M{V}^\herm$
\begin{equation}
\label{eq:CMred2}
\M{V}^\herm \M{X} \M{V}\M{V}^\herm \Ivec_n = \lambda_n  \M{\Lambda}^\herm \M{\Lambda} \M{V}^\herm \Ivec_n.
\end{equation}
Formulation~\eqref{eq:CMred2} can formally be expressed as a \ac{GEP} with an already diagonalized right hand side \cite{AngiulliVenneri_SDTforCM}
\begin{equation}
\label{eq:CMred3}
\widetilde{\M{X}} \widetilde{\Ivec}_n = \lambda_n  \widetilde{\M{R}} \widetilde{\Ivec}_n,
\end{equation}
\ie{}, $\widetilde{\M{X}}\equiv \M{V}^\herm \M{X} \M{V}$, $\widetilde{\M{R}}\equiv \M{\Lambda}^\herm \M{\Lambda}$, and $\widetilde{\Ivec}_n \equiv \M{V}^\herm \Ivec_n$. 

Since the matrix~$\Smat$ is in general rectangular, it is crucial to take into account cases where \mbox{$\Nu<\Npsi$},~\eqref{eq:numberOfSphWaves}. This is equivalent to a situation in which there are limited number of spherical projections to recover the \acp{CM}. Consequently, only limited number of singular values $\Lambda_{nn}$~exist. In such a case, the procedure similar to the one used in~\cite{HarringtonMautz_ComputationOfCharacteristicModesForConductingBodies} should be undertaken by partitioning~\eqref{eq:CMred3} into two linear systems 
\begin{equation}
\label{eq:CMred4}
\widetilde{\M{X}}\widetilde{\Ivec}=
\begin{pmatrix} 
\widetilde{\M{X}}_{11} & \widetilde{\M{X}}_{12} \\ 
\widetilde{\M{X}}_{21} & \widetilde{\M{X}}_{22}
\end{pmatrix}
\begin{pmatrix} 
\widetilde{\Ivec}_{1n} \\ 
\widetilde{\Ivec}_{2n} 
\end{pmatrix}
= 
\begin{pmatrix} 
\lambda_{1n} \widetilde{\M{R}}_{11} \widetilde{\Ivec}_{1n} \\ 
\M{0} 
\end{pmatrix},
\end{equation}
where \mbox{$\widetilde{\Ivec}_{1n} \in \mathbb{C}^{\Nu}$}, \mbox{$\widetilde{\Ivec}_{2n} \in \mathbb{C}^{\Npsi-\Nu}$}, and $\Nu<\Npsi$. The Schur complement is obtained by substituting the second row of~\eqref{eq:CMred4} into the first row
\begin{equation}
\label{eq:CMred5}
\left(\widetilde{\M{X}}_{11} - \widetilde{\M{X}}_{12} \widetilde{\M{X}}_{22}^{-1} \widetilde{\M{X}}_{21}\right) \widetilde{\Ivec}_{1n} = \lambda_{1n} \widetilde{\M{R}}_{11} \widetilde{\Ivec}_{1n}
\end{equation}
with expansion coefficients of \acp{CM} defined as
\begin{equation}
\label{eq:CMred6}
\widetilde{\Ivec}_n = \begin{pmatrix} 
\widetilde{\Ivec}_{1n} \\ 
- \widetilde{\M{X}}_{22}^{-1} \widetilde{\M{X}}_{21} \widetilde{\Ivec}_{1n}
\end{pmatrix}.
\end{equation}
As far as the matrices~$\M{U}$ and~$\M{V}$ in \eqref{eq:SVD} are unitary, the decomposition~\eqref{eq:CMred3} yields \acp{CM} implicitly normalized to
\begin{equation}
\label{eq:CMred3A}
\widetilde{\Ivec}_n^\herm \widetilde{\M{R}} \widetilde{\Ivec}_m = \delta_{nm},
\end{equation}
which is crucial since the standard normalization cannot be used without decreasing the number of significant digits. In order to demonstrate the use of \eqref{eq:CMred5}, various examples from Table~\ref{tab:Examples} are calculated and compared with the conventional approach~\eqref{eq:CMred1}. 

\begin{figure}[t]
	\begin{center}
		\includegraphics[width=\figwidth cm]{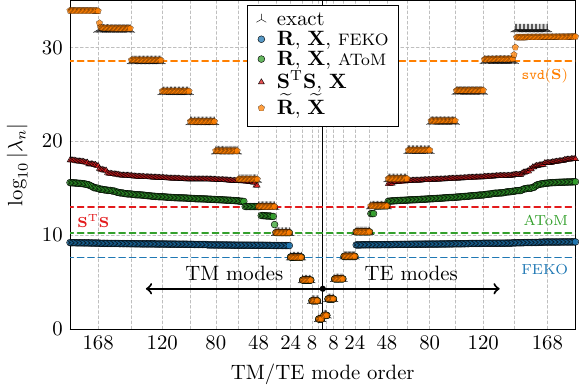}
		\caption{The absolute values of the \acp{CM} of spherical shell with electrical size $ka=0.5$ (Example~\hyperref[tab:Examples]{S2}). Data calculated with classical procedure \eqref{eq:CMred1} are compared with techniques from this paper, \eqref{eq:CMred2}, \eqref{eq:CMred5}, and with the analytical results valid for the spherical shell~\cite{CapekEtAl_ValidatingCMsolvers}. }
		\label{fig:CMdecomInit_Sphere750}
	\end{center}
\end{figure}

\begin{figure}[t]
	\begin{center}
		\includegraphics[width=\figwidth cm]{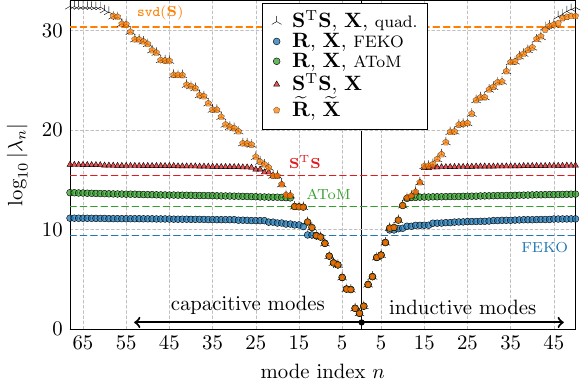}
		\caption{The absolute values of the \acp{CM} of rectangular plate (Example~\hyperref[tab:Examples]{R2}). Since unknown analytical results, the multiprecision package Advanpix has been used instead to calculate the first $150$~modes from impedance matrix in quadruple precision.}
		\label{fig:CMdecomInit_Rectangle655} 
	\end{center}
\end{figure}

The \acp{CM} of the spherical shell from Example~\hyperref[tab:Examples]{S2} are calculated and shown as absolute values in logarithmic scale in Fig.~\ref{fig:CMdecomInit_Sphere750}. It is shown that the number of the \acp{CM} calculated by classical procedure (FEKO, AToM) is limited to the lower modes, especially considering the degeneracy \mbox{$2l+1$} of the \acp{CM} on the spherical shell~\cite{CapekEtAl_ValidatingCMsolvers}. The number of properly found \acp{CM} is significantly higher when using \eqref{eq:CMred5} than the conventional approach \eqref{eq:CMred1} and the numerical dynamic is doubled. Notice that, even \eqref{eq:CMred1} where the matrix~$\M{R}$ calculated from~\eqref{eq:Rdecomp} yields slightly better results than the conventional procedure. This fact is confirmed in Fig.~\ref{fig:CMdecomInit_Rectangle655} dealing with Example~\hyperref[tab:Examples]{\hyperref[tab:Examples]{R2}}, where the multiprecision package Advanpix is used as a reference. The same calculation illustrates that the matrix~$\M{R}$ contains all information to recover the same number of modes as \eqref{eq:CMred5}, but this can be done only at the expense of higher computation time\footnote{For Example~\hyperref[tab:Examples]{S2} the computation time of \acp{CM} with quadruple precision is approximately $15$~hours.}.

\begin{figure}[t]
	\begin{center}
		\includegraphics[width=\figwidth cm]{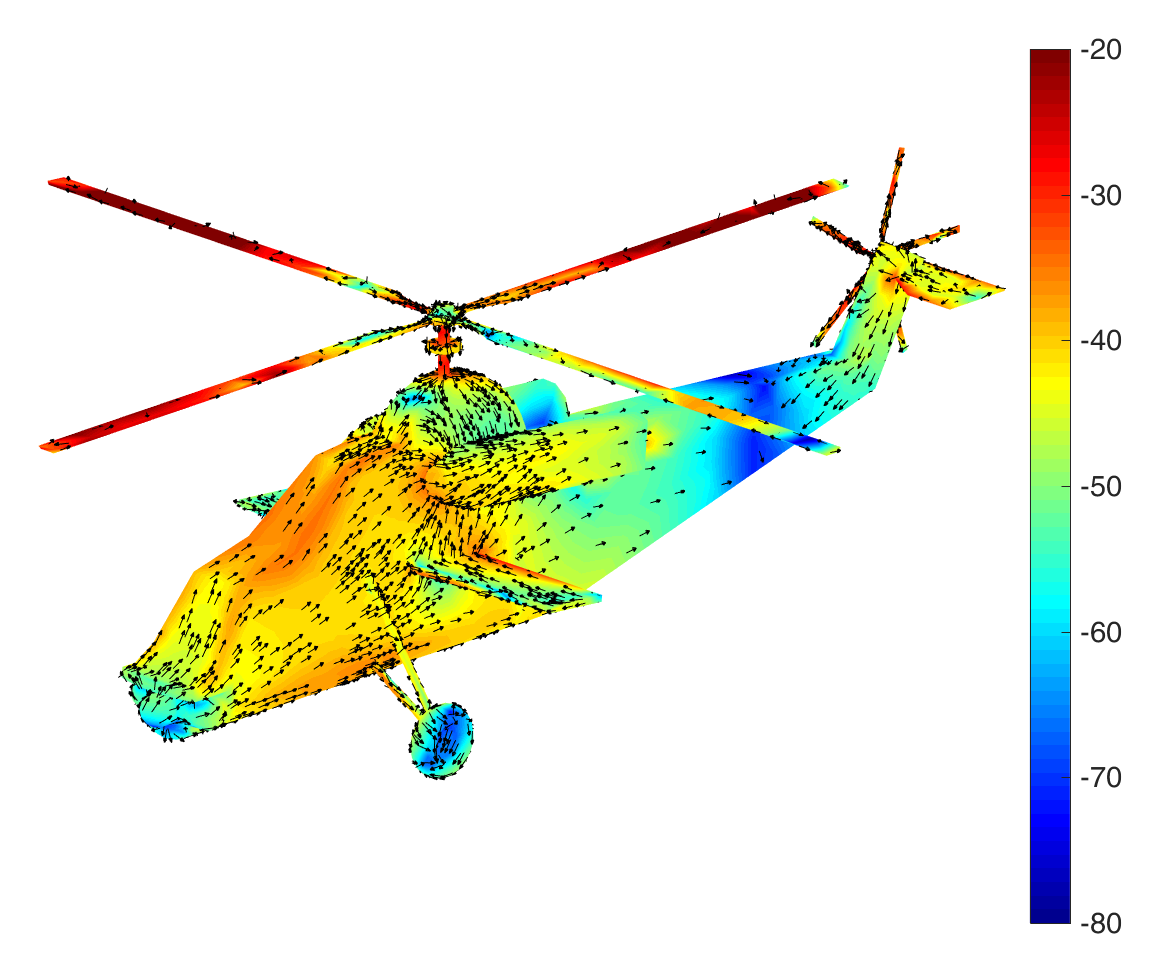}
		\caption{\TRa{Current density of the} first characteristic mode of a helicopter \TRa{at $ka=7$} (Example~\hyperref[tab:Examples]{H2}), \TRa{mesh grid has been taken from}~\cite{feko}.}
		\label{fig:CMdecomHeli}
	\end{center}
\end{figure}

While ~\eqref{eq:CMred5} preserves the numerical dynamics, the computational efficiency is not improved due to the matrix multiplications to calculate the $\widetilde{\M{X}}$ term in~\eqref{eq:CMred4}. An alternative formulation that improves the computational speed is derived by replacing the matrix $\M{R}$ with~\eqref{eq:Rdecomp} in~\eqref{eq:CMred1}

\begin{equation}
\label{eq:SXS1}
\M{X} \Ivec_n = \lambda_n \Smat^\trans \Smat \Ivec_n,
\end{equation}
and multiplying from the left with $\Smat \M{X}^{-1}$
\begin{equation}
\label{eq:SXS2}
\Smat \Ivec_n = \lambda_n \Smat \M{X}^{-1} \Smat^\trans \Smat \Ivec_n.
\end{equation}
The formulation~\eqref{eq:SXS2} is a standard eigenvalue problem and can be written as
\begin{equation}
\label{eq:SXS3} 
\Smat \M{X}^{-1} \Smat^\trans \widehat{\Ivec}_n= \widehat{\M{X}} \widehat{\Ivec}_n = \xi_n \widehat{\Ivec}_n,
\end{equation}
where 
$\widehat{\M{X}} = \Smat \M{X}^{-1} \Smat^\trans$, $\widehat{\Ivec}_n = \Smat \Ivec$, and $\xi_n = 1/\lambda_n$. As an intermediary step, the matrix $\M{X}_\mathrm{S}=\M{X}^{-1}\Smat^\trans$ is computed, which is later used to calculate the characteristic eigenvectors $\Ivec_n=\lambda_n\M{X}_\mathrm{S}\widehat{\Ivec}_n$. The eigenvalue problem~\eqref{eq:SXS3} is solved in the basis of spherical vector waves, $\widehat{\Ivec}_n = \Smat \Ivec$, that results in a matrix $\widehat{\M{X}} \in \mathbb{C}^{\Nu\times\Nu}$. For problems with $\Nu\ll\Npsi$ the eigenvalue problem is solved rapidly compared with~\eqref{eq:CMred1} and~\eqref{eq:CMred5}. The computation times for various examples are presented in Table~\ref{tab:CMcompTimes} for all three formulations where a different number of \acp{CM} are compared. For Example~\hyperref[tab:Examples]{H1} the computation time is investigated for the first $20$ and $100$ modes. The acceleration using~\eqref{eq:SXS3} is approximately $4.7$ and $14$ times when compared with the conventional method~\eqref{eq:CMred1}. The first characteristic mode of Example~\hyperref[tab:Examples]{H1} is illustrated in Fig.~\ref{fig:CMdecomHeli}.

\begin{table}[t]
	\begin{center}
		\begin{tabular}{cccccc}
			\toprule
			Example & & \multicolumn{3}{c}{Time to calculate $N_\lambda$ \acp{CM} (s)} \\ 
			\cmidrule{3-5}
			(see Table~\ref{tab:Examples}) & $N_\lambda$ & \eqref{eq:CMred1} & \eqref{eq:CMred5} & \eqref{eq:SXS3}  \\ \toprule
			\hyperref[tab:Examples]{S1} & $10$ & $0.36$ & $0.18$ & $0.12$  \\             
			\hyperref[tab:Examples]{S2} & $300$ & $3.3$ & $2.0$ & $1.1$  \\ 
			\hyperref[tab:Examples]{S4} & $10$ & $2.8$ & $2.5$ & $0.78$  \\             
			\hyperref[tab:Examples]{S4} & $100$ & $13$ & $2.1$ & $0.72$ \\ \midrule                        
			\hyperref[tab:Examples]{R1} & $100$ & $0.29$ & $0.28$ & $0.42$  \\ 
			\hyperref[tab:Examples]{R3} & $50$ & $7.2$ & $1.3$ & $0.49$  \\\midrule
			\hyperref[tab:Examples]{H1} & $20$ & $130$ & $150$ & $28$  \\ 
			\hyperref[tab:Examples]{H1} & $100$ & $500$ & $150$ & $35$  \\ \iftoggle{ARXIV}{}{} 
			\hyperref[tab:Examples]{H2} & $100$ & $350$ & $160$ & $35$  \\                              
			\bottomrule
		\end{tabular} 
	\end{center}
	\caption{Comparison of computation time required by various methods capable to calculate first $N_\lambda$ CMs. 
	The calculations were done on Windows~Server~2012 with 2$\times$Xeon E5-2665 CPU @ $2.4$~GHz and $72$~GB RAM.}
	\label{tab:CMcompTimes}
\end{table}

\begin{figure}[t]
	\begin{center}
		\includegraphics[width=\figwidth cm]{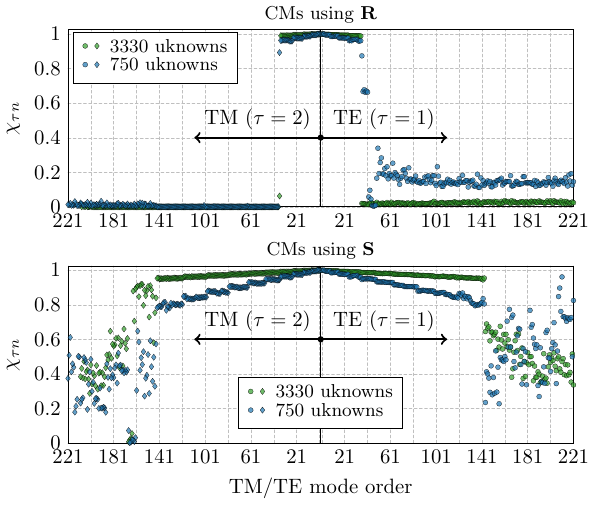}
		\caption{Similarity of numerically evaluated characteristic currents for a spherical shell of two different discretizations (Example~\hyperref[tab:Examples]{S2} and~\hyperref[tab:Examples]{S5}) and the analytically known currents \cite{CapekEtAl_ValidatingCMsolvers}. The coefficients~$\chi_{\tau n}$ were calculated according to~\cite{CapekEtAl_ValidatingCMsolvers}, top panel depicts results for the conventional procedure~\eqref{eq:CMred1}, bottom panel for the procedure from this paper~\eqref{eq:CMred5}.}
		\label{fig:CMdecomSpher_JtestComp}
	\end{center}
\end{figure}

Two tests proposed in~\cite{CapekEtAl_ValidatingCMsolvers} are performed to validate the conformity of characteristic current densities and the characteristic far fields with the analytically known values. The results of the former test are depicted in Fig.~\ref{fig:CMdecomSpher_JtestComp} for Example~\hyperref[tab:Examples]{S2} and~\hyperref[tab:Examples]{S5} that are spherical shells with two different \ac{dof}. Similarity coefficients~$\chi_{\tau n}$ are depicted both for the \acp{CM} using the matrix~$\M{R}$ \eqref{eq:CMred1} and for the \acp{CM} calculated by~\eqref{eq:CMred5}. The number of valid modes correlates well with Table~\ref{tab:FoundModes} and the same dependence on the quality and size of the mesh grid as in~\cite{CapekEtAl_ValidatingCMsolvers} is observed. 

\begin{figure}[t]
	\begin{center}
		\includegraphics[width=\figwidth cm]{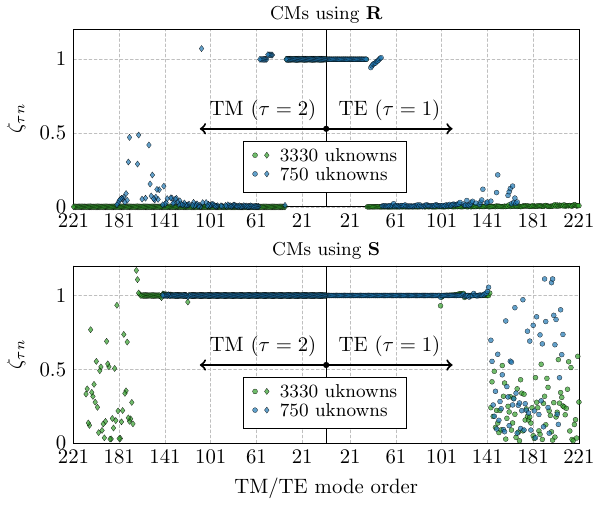}
		\caption{Similarity of numerically and analytically evaluated characteristic far fields for a spherical shell of two different discretizations (Example~\hyperref[tab:Examples]{S2} and~\hyperref[tab:Examples]{S5}) and analytically known far fields \cite{CapekEtAl_ValidatingCMsolvers}. The coefficients~$\zeta_{\tau n}$ were calculated \TRa{by \eqref{eq:Ftest1}} \TRsa{according to} \TRa{, see} \cite{CapekEtAl_ValidatingCMsolvers} \TRa{for more details.} \TRsa{, top} \TRa{Top} panel depicts results for the conventional procedure~\eqref{eq:CMred1}, bottom panel for the procedure from this paper~\eqref{eq:CMred5}.}
		\label{fig:CMdecomSpher_FtestComp}
	\end{center}
\end{figure}

Qualitatively the same behavior is also observed in the latter test, depicted in Fig.~\ref{fig:CMdecomSpher_FtestComp}, where similarity of characteristic far fields is expressed by coefficient~$\zeta_{\tau n}$~\mbox{\cite{CapekEtAl_ValidatingCMsolvers}}. These coefficients read
\TRa{
\begin{equation}
\label{eq:Ftest1}
\zeta_{\tau n} = \max\limits_{l}\sum_{\sigma m}\left| \tilde{f}_{\tau\sigma mln}\right|^2,
\end{equation}
}where \TRa{$\tilde{f}_{\tau\sigma mln}$ has been evaluated using~\eqref{eq:Ftest4}.}
\TRsa{with \mbox{$\V{F}_n$} being the characteristic far fields evaluated for a spherical shell using~\eqref{eq:Ftest3} with $[\widetilde{f}_{\alpha}] = \Smat \Ivec_n$.}
The \TRa{results for} characteristic far fields computed from the conventional procedure~\eqref{eq:CMred1} and the procedure presented in this paper~\eqref{eq:CMred5} are illustrated in Fig.~\ref{fig:CMdecomSpher_FtestComp}. 

Lastly, the improved accuracy of using~\eqref{eq:CMred5} over~\eqref{eq:CMred1}, is demonstrated in the Fig.~\ref{fig:RectModes1} \TRsa{, for the $17^{\mathrm{th}}$ inductive \ac{CM} of the rectangular plate (Example~\hyperref[tab:Examples]{R2}). The surface current density in the left panel, calculated using $\mathbf{I}_{41}$ in \eqref{eq:CMred1} is, in fact, only the numerical noise. However, in the right panel, the current density calculated using \eqref{eq:CMred5} is the correct higher-order mode.} \TRa{which shows current profiles, corresponding to a rectangular plate (Example~\hyperref[tab:Examples]{R2}),
\iftoggle{ARXIV}{%
of a collection of the first 30 modes.
}
{of a selected high order mode (a collection of the first 30 modes is presented in~\cite{TayliEtAl_AccurateAndEfficientEvaluationofCMs}).}
It can be seen that for modes with high eigenvalues (numerically saturated regions in~Fig.~\ref{fig:CMdecomInit_Rectangle655}) the surface current density in left panel, calculated via~\eqref{eq:CMred1}, shows numerical noise, while the evaluation via~\eqref{eq:CMred5} still yields a correct current profile.}
\iftoggle{ARXIV}{%
\begin{figure}[t]
	\centering
	\resizebox{4.3 cm}{!}{\animategraphics[loop, controls, autoplay, buttonsize=0.5em, poster=28]{1}{./TikZ_animate/OLD_merge}{0}{30}}
	\resizebox{4.3 cm}{!}{\animategraphics[loop, controls, autoplay, buttonsize=0.5em, poster=28]{1}{./TikZ_animate/NEW_merge}{0}{30}}
	\caption{Comparison of the \TRsa{$17^\mathrm{th}$ inductive} \TRa{higher-order \acp{CM}} \TRsa{\ac{CM}} of the rectangular plate (Example~\hyperref[tab:Examples]{R2}) \TRa{with the most similar characteristic number}, left panel: conventional procedure \eqref{eq:CMred1}, right panel: procedure from this paper \eqref{eq:CMred5}.}
	\label{fig:RectModes1}
\end{figure}
}
{
\begin{figure}[t]
	\centering
	\resizebox{4.3 cm}{!}{\includegraphics{Fig9_Rectangle655_Modes_FigA.pdf}}%
	\resizebox{0.1 cm}{!}{\,}	
	\resizebox{4.3 cm}{!}{\includegraphics{Fig9_Rectangle655_Modes_FigB.pdf}}%
	\caption{Comparison of the \TRsa{$17^\mathrm{th}$ inductive} \TRa{higher-order \acp{CM}} \TRsa{\ac{CM}} of the rectangular plate (Example~\hyperref[tab:Examples]{R2}) \TRa{with the most similar characteristic number}, left panel: conventional procedure \eqref{eq:CMred1}, right panel: procedure from this paper \eqref{eq:CMred5}. \TRa{The first 30 modes evaluated via both procedures are available as interactive collection (available in Adobe Acrobat Reader) in~\cite{TayliEtAl_AccurateAndEfficientEvaluationofCMs}, \cf{} Fig.~\ref{fig:CMdecomInit_Rectangle655}.}}
	\label{fig:RectModes1}
\end{figure}
}

\subsection{Restriction to TM/TE modes}
\label{sec:TMTE}

Matrix~$\Smat$, described in Section~\ref{sec:SPH}, contains projections onto TE and TM spherical waves in its odd ($\tau=1$) and even rows ($\tau=2$), respectively. The separation of TE and TM spherical waves can be used to construct resistance matrices~$\M{R}^\mathrm{TE}$ and $\M{R}^\mathrm{TM}$, where only odd and even rows of matrix~$\Smat$ are used to evaluate \eqref{eq:Rdecomp}.

\begin{figure}[t]
	\centering
	\includegraphics[width=\figwidth cm]{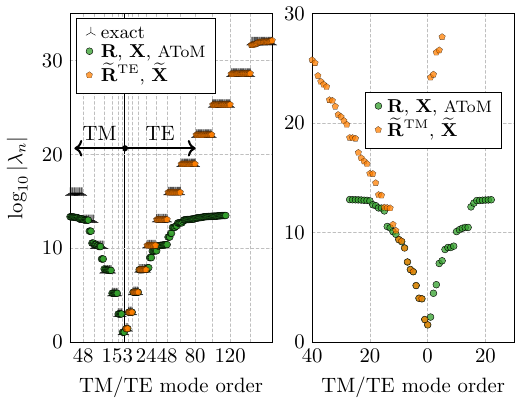}
	\caption{Left pane: the absolute values of the CMs of a spherical shell (Example~\hyperref[tab:Examples]{S2}) if only odd rows of the matrix~$\Smat$ were kept. Right pane: the absolute values of the CMs of a rectangular plate (Example~\hyperref[tab:Examples]{R2}) if only even rows of the matrix~$\Smat$ were kept.}
	\label{fig:STMTErestriction}
\end{figure}

Matrices $\M{R}^\mathrm{TM}$ and $\M{R}^\mathrm{TE}$ can be used in optimization, \eg{}, in such a case when the antennas have to radiate TM-modes only~\cite{Capek+etal2017b}. With this feature, characteristic modes consisting of only TM (or TE) modes can easily be found. This is shown in Fig.~\ref{fig:STMTErestriction}, in which the spherical shell (Example~\hyperref[tab:Examples]{S2}) and rectangular plate (Example~\hyperref[tab:Examples]{R2}) are used to find only TM (capacitive) and TE (inductive) modes, respectively. In case of a spherical shell this separation could have been done during the post-processing. For a generally shaped body this separation however represents a unique feature of the proposed method.

\section{Discussion}
\label{sec:Disc}

Important aspects of the utilization of the matrix~$\Smat$ are discussed under the headings implementation aspects, computational aspects and potential improvements.

\subsection{Implementation Aspects}
\label{sec:Dis1p5}

Unlike the reactance matrix~$\M{X}$, the resistance matrix~$\M{R}$ suffers from high condition number.
Therefore, the combined approach to evaluate the impedance matrix (matrix~$\M{R}$ using matrix~$\Smat$, matrix~$\M{X}$ using conventional Green function technique with double integration) takes advantage of both methods and is optimal for, \eg{}, modal decomposition techniques dealing with the matrix~$\M{R}$ (radiation modes~\mbox{\cite{SchabBernhard_RadiationAndEnergyStorageCurrentModesOnConductingStructures}}, \acp{CM}, energy modes~\mbox{\cite{SchabBernhard_RadiationAndEnergyStorageCurrentModesOnConductingStructures, JelinekCapek_OptimalCurrentsOnArbitrarilyShapedSurfaces}}, and solution of optimization problems~\mbox{\cite{Capek+etal2017b}}). Evaluation and the \ac{SVD} of the matrix~$\Smat$ are also used to estimate number of modes, \cf{} number of modes of the matrix~$\Smat$ found by~\eqref{eq:SVD} and number of \acp{CM} found by~\eqref{eq:CMred5} in Table~\ref{tab:FoundModes}.


\subsection{Computational Aspects}
\label{sec:Dis2}


Computational gains of the proposed method are seen in Table~\ref{tab:Speed} for the matrix $\M{R}$ and Table~\ref{tab:CMcompTimes} for the \acp{CM}. The formulation~\eqref{eq:SXS3} significantly accelerates \acp{CM} computation when compared with the classical \ac{GEP} formulation~\eqref{eq:CMred1}. Moreover, it is possible to employ lower precision floating point arithmetic, \eg{} float, to compute as many modes as the conventional method that employs higher precision floating point arithmetic, \eg{} double. In modern hardware, this can provide additional performance boosts if vectorization is used.


An advantage of the proposed method is that the matrix~$\Smat$ is rectangular for $\Nu<\Npsi$, allowing independent selection of the parameters $\Npsi$ and $\Nu$. While the parameter~$\Npsi$ controls the details in the model, the parameter $\Nu$ (or alternatively $L$) controls the convergence of the matrix~$\Smat$ and the number of modes to be found.
In this paper~\eqref{eq:Lmax} is used to determine the highest spherical wave order~$L$ for a given electrical size $ka$. The parameter~$L$ can be increased for improved accuracy or decreased for computational gain depending on the requirements of the problem. Notice that the parameter $\Nu$ is limited from below by the convergence and the number of desired modes, but also from above since the spherical Bessel function in~$\UFCN{\alpha}{1}{k\V{r}}$ decays rapidly with $l$ as
\begin{equation}
\label{eq:sphBessLim}
\mathrm{j}_l \left(ka\right) \approx \frac{2^l l!}{\left(2l + 1\right)!} \left(ka\right)^l,\quad  ka\ll l.
\end{equation}
The rapid decay can be observed in Fig.~\ref{fig:Plate_Lstudy}, where the convergence of the matrix $\M{R}$ to double precision for $ka=3$ requires only $L=12$ while (16) gives a conservative number of $L=17$.
\subsection{Potential Improvements}
\label{sec:Dis3}

Even though the numerical dynamic is increased, it is strictly limited and it presents an inevitable, thus fundamental, bottleneck of all modal methods involving radiation properties. 
The true technical limitation is, in fact, the \ac{SVD} of the matrix~$\Smat$. A possible remedy is the use of high-precision packages that come at the expense of markedly longer computation times and the necessity of performing all subsequent operations in the same package to preserve high numerical precision.

The second potential improvement relies on higher-order basis functions, which can compensate a poor-meshing scheme (that is sometimes unavoidable for complex or electrically large models). It can also reduce the number of basis function~$\Npsi$ so that the evaluation of \acp{CM} is further accelerated.

\section{Conclusion}
\label{sec:Concl}

Evaluation of the discretized form of the EFIE impedance operator, the impedance matrix, has been reformulated using projection of vector spherical harmonics onto a set of basis functions. The key feature of the proposed method is the fact that the real part of the impedance matrix can be written as a multiplication of the spherical modes projection matrix with itself. This feature accelerates modal decomposition techniques and doubles the achievable numerical dynamics. The results obtained by the method can also be used as a reference for validation and benchmarking.



It has been shown that the method has notable advantages, namely the number of available modes can be estimated prior to the decomposition and the convergence can be controlled via the number of basis functions and the number of projections. The normalization of generalized eigenvalue problems with respect to the product of the spherical modes projection matrix on the right hand side are implicitly done. The presented procedure finds its use in various optimization techniques as well. It allows for example to prescribe the radiation pattern of optimized current by restricting the set of the spherical harmonics used for construction of the matrix.

The method can be straightforwardly implemented into both in-house and commercial solvers, improving thus their performance and providing antenna designers with more accurate and larger sets of modes.

\appendices

\section{Used Computational Electromagnetics Packages}
\label{app:CEMpackages}
\subsection{FEKO}
FEKO (ver.~14.0-273612,~\cite{feko}) has been used with a mesh structure that was imported in NASTRAN file format \cite{nastran}: CMs and far fields were chosen from the model tree under \emph{requests} for the FEKO solver. Data from FEKO were acquired using *.out, *.os, *.mat and *.ffe files. The impedance matrices were imported using an in-house wrapper \cite{IDA}. Double precision was enabled for data storage in solver settings.

\subsection{AToM}
AToM (pre-product ver., CTU in Prague,~\cite{atom}) has been used with a mesh grid that was imported in NASTRAN file format \cite{nastran}, and simulation parameters were set to comply with the data in Table~\ref{tab:Examples}. AToM uses RWG basis functions with the Galerkin procedure~\cite{RaoWiltonGlisson_ElectromagneticScatteringBySurfacesOfArbitraryShape}. The Gaussian quadrature is implemented according to~\cite{Dunavant_HighDegreeEfficientGQR} and singularity treatment is implemented from~\cite{EibertHansen_OnTheCalculationOfPotenticalIntegralsForLinearSourceDistributionsOnTriangularDomains}. Built-in Matlab functions are utilized for matrix inversion and decomposition. Multiprecision package Advanpix \cite{advanpix} is used for comparison purposes.

\subsection{IDA}
IDA (in-house, Lund University,~\cite{IDA}) has been used with the NASTRAN mesh and processed with the IDA geometry interpreter. IDA solver is a Galerkin type \ac{MoM} implementation. RWG basis functions are used for the current densities. Numerical integrals are performed using Gaussian quadrature~\cite{Dunavant_HighDegreeEfficientGQR} for \mbox{non-singular} terms and the DEMCEM library~\cite{DEMCEM2010,Polimeridis+Yioultsis2008,Polimeridis+Wilton2010,Polimeridis+Mosig2011} for singular terms. Intel MKL library~\cite{IntelMKL2017} is used for linear algebra routines. The matrix computation routines are parallelized using OpenMP~$2.0$~\cite{Dagum1998}. Multiprecision computations were done with the mpmath Python library~\cite{mpmath2013}.

\section{Spherical Vector Waves}
\label{app:SPHwaves}

General expression of the (scalar) spherical modes is~\cite{Kristensson2016}
\begin{equation}
\label{eq:sphWave1}
\mathrm{u}^{(p)}_{\sigma ml}(k\V{r})= \mathrm{z}_{l}^{(p)}(kr)\YFCNs{\sigma m l}{\UV{r}},
\end{equation}
with \mbox{$\UV{r}=\V{r}/|\V{r}|$} and $k$ being the wavenumber. The indices are \mbox{$m \in \left\{ 0,\dots,l \right\}$}, \mbox{$\sigma \in \left\{ \mathrm{e},\mathrm{o} \right\}$} and \mbox{$l \in \left\{ 1,\dots,L \right\}$}~\cite{Hansen1988,Gustafsson+Nordebo2006c}. For regular waves \mbox{$\mathrm{z}_l^{(1)}=\mathrm{j}_l$} is a spherical Bessel function of order $l$, irregular waves \mbox{$\mathrm{z}_l^{(2)}=\mathrm{n}_l$} is a spherical Neumann function, and  \mbox{$\mathrm{z}_l^{(3,4)}=\mathrm{h}^{(1,2)}_l$} are spherical Hankel functions for the ingoing and outgoing waves, respectively. Spherical harmonics are defined as~\cite{Kristensson2016}
\begin{equation}
\YFCNs{\sigma m l}{\UV{r}} = \sqrt{\frac{\varepsilon_m}{2\pi}}\widetilde{P}_l^m \left(\cos\vartheta\right)
\begin{Bmatrix}
\cos m\varphi\\
\sin m\varphi
\end{Bmatrix}
,\quad 
\sigma
=
\begin{Bmatrix}
\mathrm{e}\\
\mathrm{o}
\end{Bmatrix}
\end{equation}
with $\varepsilon_m=2-\delta_{m0}$ the Neumann factor, $\delta_{ij}$ the Kronecker delta function and $\PFCNn{l}{m}{\cos\vartheta}$ the normalized associated Legendre functions~\cite{Olver+etal2010}. 

The spherical vector waves are~\cite{Hansen1988, Kristensson2016}
\begin{subequations}
\begin{align}
\label{eq:sphWaveU1}
\UFCN{1\sigma  m l}{p}{k\V{r}} &= \RFCN{1 l}{p}{kr}\YFCN{1\sigma m l}{\UV{r}}, \\
\label{eq:sphWaveU2}
\UFCN{2\sigma  m l}{p}{k\V{r}} &= \RFCN{2 l}{p}{kr}\YFCN{2\sigma m l}{\UV{r}} + \RFCN{3 l}{p}{k r} \YFCNs{\sigma m l}{\UV{r}} \UV{r},
\end{align}
\end{subequations}
where $\RFCN{\tau l}{p}{kr}$ are the radial function of order $l$ defined as
\begin{subnumcases}
{\RFCN{\tau l}{p}{\kappa}=}	
\label{eq:Rfunc1}
\mathrm{z}_l^{(p)}(\kappa), & $\tau=1$, \\
\label{eq:Rfunc2}
\frac{1}{\kappa}\frac{\partial}{\partial\kappa}\left(\kappa \mathrm{z}_l^{(p)}\left(\kappa\right)\right), & $\tau=2$, \\
\label{eq:Rfunc3}
\frac{\ConstB}{\kappa}\mathrm{z}_l^{(p)}(\kappa), & $\tau=3$,
\end{subnumcases}
with $\ConstB = \sqrt{l \left( l+1 \right)}$ and \mbox{$\YFCN{\tau\sigma  m l}{\UV{r}}$} denotes the \mbox{real-valued} vector spherical harmonics defined as 
\begin{subequations}
\begin{align}
\label{eq:Ycomp1}
\YFCN{1\sigma m l}{\UV{r}}	&= \frac{1}{\ConstB}\nabla\times\left(\V{r} \YFCNs{\sigma m l}{\UV{r}}\right), \\
\label{eq:Ycomp2}
\YFCN{2\sigma m l}{\UV{r}} &= \UV{r} \times \YFCN{1\sigma m l}{\UV{r}},
\end{align}
\end{subequations}
where $\mathrm{Y}_{\sigma m l}$ denotes the ordinary spherical harmonics~\cite{Kristensson2016}. The radial functions can be seperated into real and imaginary parts as
\begin{align}
\RFCN{\tau l}{3}{\kappa} &= \RFCN{\tau l}{1}{\kappa} +\J \RFCN{\tau l}{2}{\kappa},\\
\RFCN{\tau l}{4}{\kappa} &= \RFCN{\tau l}{1}{\kappa} -\J \RFCN{\tau l}{2}{\kappa}.
\end{align}

\section{Associated Legendre Polynomials}
\label{app:DerivALP}

The associated Legendre functions are defined \cite{Jeffrey_MathHandbook} as
\begin{equation}
\label{eq:AssocLegPol1}
P_l^m \left( x \right) = \left( 1-x^2 \right)^{m/2} \frac{\D{}^m }{\D{} x^m}P_{l}(x), \quad l\geq m \geq 0,
\end{equation}
with
\begin{equation}
\label{eq:AssocLegPol2}
P_l \left( x \right) = \frac {1}{2^{l}l!}\frac {\D{}^{l}}{\D{} x^l}\left(x^2 - 1 \right)^l
\end{equation}
being the associated Legendre polynomials of degree~$l$ and \mbox{$x \in \left[-1, 1 \right]$}. One useful limit when computing the vector spherical harmonics is~\cite{Kristensson2016}
\begin{equation}
\lim_{x\to 1} \frac{P_l^m \left( x \right)}{\sqrt{1-x^2}}=\delta_{m1}\frac{l\left(l+1\right)}{2}.
\end{equation}
The normalized associated Legendre function $\tilde{P}_{l}^{m}$, is defined as follows
\begin{equation}
\label{eq:NALF}
\widetilde{P}_l^m \left(x\right) = \sqrt{\frac{2l+1}{2}\frac{(l-m)!}{(l+m)!}} P_l^m \left(x\right).
\end{equation}
The derivative of the normalized associated Legendre function is required when computing the spherical harmonics, and is given by the following recursion relation
\begin{multline}
\label{eq:DerivAssocLegPol1}
\frac{\partial }{\partial \vartheta}\widetilde{P}_l^m \left( \cos\vartheta \right)  = \frac{1}{2}\sqrt{(l+m)(l-m+1)}\widetilde{P}_l^{m-1} \left( \cos\vartheta \right)\\
-\frac{1}{2}\sqrt{(l-m)(l+m+1)}\widetilde{P}_l^{m+1} \left( \cos\vartheta \right)
\end{multline}
where $x\equiv \cos\vartheta$, \mbox{$\vartheta \in \left[0, \pi \right]$}.

\iftoggle{ARXIV}{%
\section{Spherical Shell and Rectangular Plate}
\label{app:Mesh}

Meshes for the spherical shell of radius $a=1\unit{m}$ with $\Npsi=750$ and $\Npsi=3330$ \ac{dof} are depicted in Fig.~\ref{fig:SphereNumericalMesh}. The meshes for the rectangular plate of aspect ratio $L/W=2$ with $\Npsi=199$ , $\Npsi=655$, and $\Npsi=2657$ \ac{dof} are presented in Fig.~\ref{fig:RectangleNumericalMesh}.
\label{app:RadMode}
\begin{figure}[t]
	\begin{center}
		\includegraphics[width=\figwidth cm]{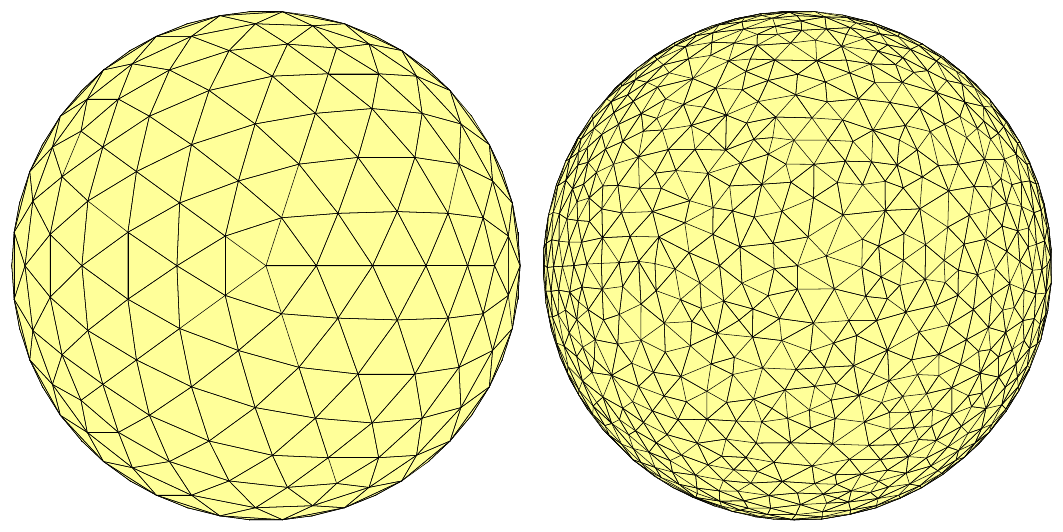}
		\caption{Spherical shell mesh with $500$~triangles (left) and $2220$~triangles (right) with $750$~(left) and $3330$~(right) RWG basis functions, respectively. The same mesh grids are used in~\cite{CapekEtAl_ValidatingCMsolvers} to make the results comparable.}
		\label{fig:SphereNumericalMesh}
	\end{center}
\end{figure}
\begin{figure}[t]
	\begin{center}
		\includegraphics[width=\figwidth cm]{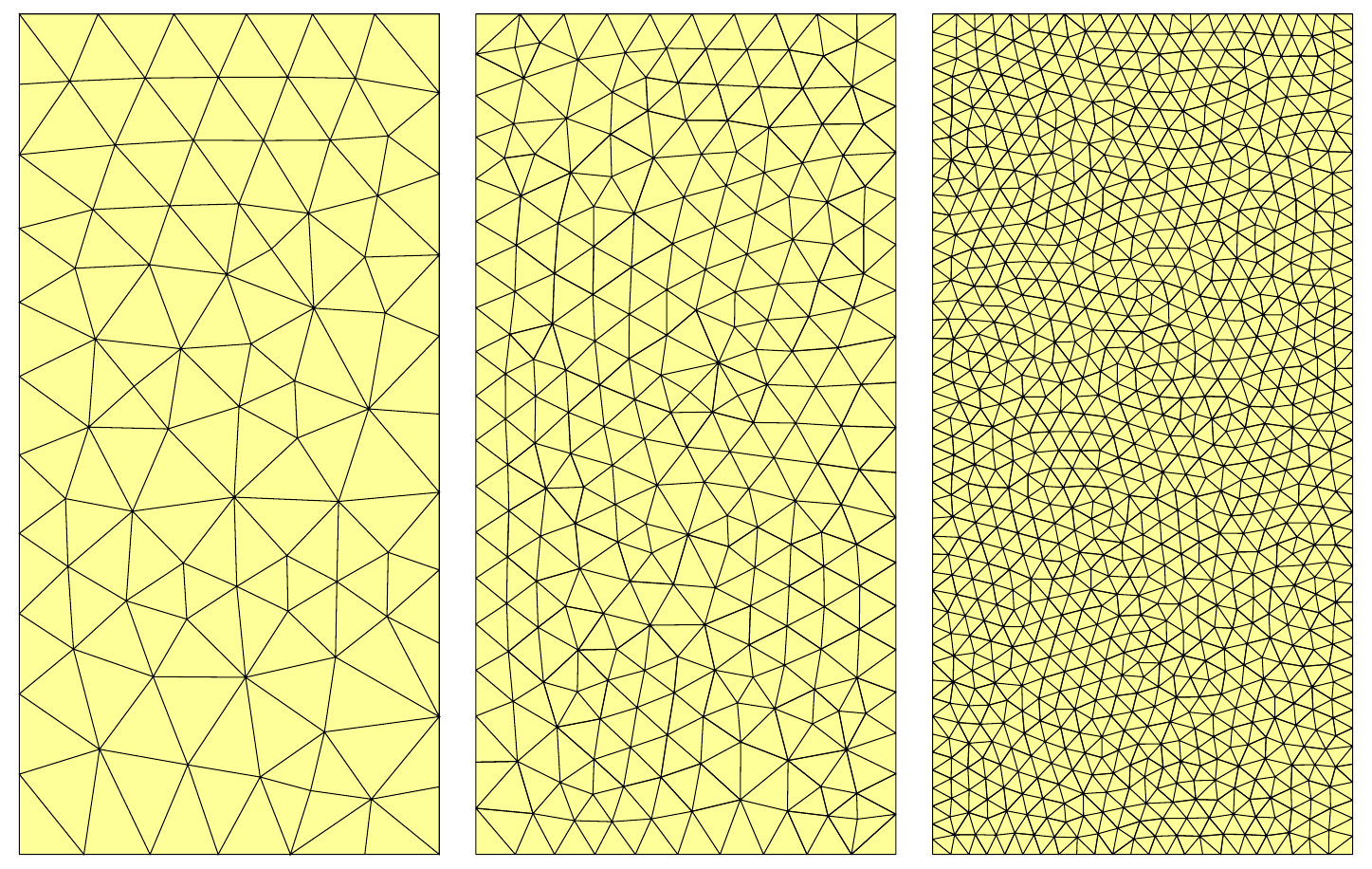}
		\caption{Rectangular plate mesh with $144$, $456$, and $1818$~triangles (from left to right) with $199$, $655$, and $2657$~RWG basis functions, respectively.}
		\label{fig:RectangleNumericalMesh}
	\end{center}
\end{figure}
}
{}
\iftoggle{ARXIV}{%
\section{Radiation Modes}
\label{app:RadMode}

Eigenvalues of the radiation modes for Example \hyperref[tab:Examples]{S2} and \hyperref[tab:Examples]{R2} are presented in Fig.~\ref{fig:Sphere_Rdecomp} and Fig.~\ref{fig:Rectangle_Rdecomp}. The eigenvalues are computed using both the conventional \eqref{eq:Rmodes} and the proposed \eqref{eq:Smodes} method. It can be seen that the number of modes computed using~\eqref{eq:Smodes} is significantly higher compared to~\eqref{eq:Rmodes} for both examples. Eigenvalues calculated using quadruple precision \ac{SVD} of the matrix $\Smat$ are also included. The number of correct radiation modes is shown in Table~\ref{tab:FoundModes}.

If eigenvalues $\xi_n$ of the different mesh grids are to be compared the \ac{MoM} matrices must be normalized.
The normalized matrices are \mbox{$\widehat{\M{R}} = \M{L} \M{R} \M{L}$}, \mbox{$\widehat{\boldsymbol{\xi}} = \M{L} \boldsymbol{\xi} \M{L}$}, \mbox{$\widehat{\M{S}} = \M{S} \M{L}$}, \mbox{$\widehat{\Ivec}_n=\M{L}^{-1}\Ivec_n $}, where $\M{L}$ is the diagonal matrix of basis functions' reciprocal edge lengths, \ie{}, \mbox{$L_{pp} = 1/l_p$}.

\begin{figure}[h]
	\begin{center}
		\includegraphics[width=\figwidth cm]{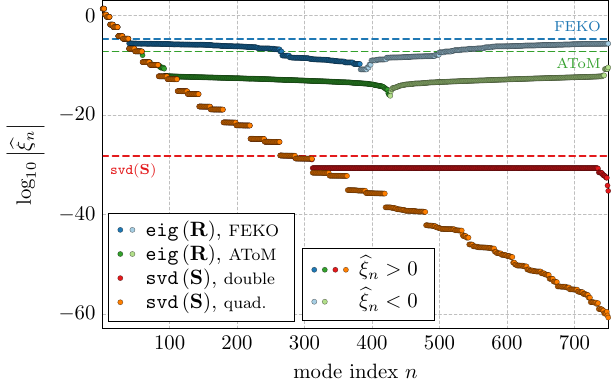}
		\caption{Normalized eigenvalues of the matrix~$\M{R}$ of a spherical shell with electrical size $ka=0.5$ discretized into $500$~triangles (Example~\hyperref[tab:Examples]{S2}). Multiprecision package Advanpix \cite{advanpix} has been used for evaluation in quadruple precision. The number of well-determined modes is delimited by horizontal dashed lines.}
		\label{fig:Sphere_Rdecomp}
	\end{center}
\end{figure}
\begin{figure}[h]
	\begin{center}
		\includegraphics[width=\figwidth cm]{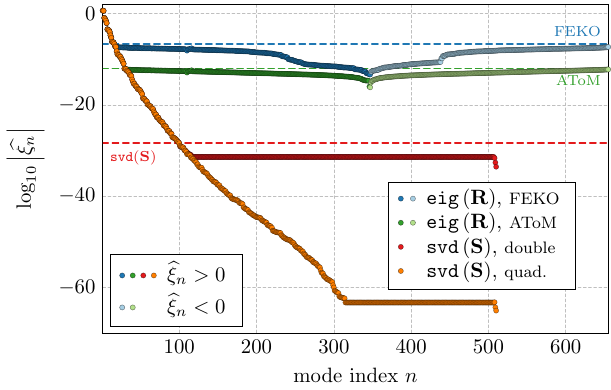}
		\caption{Normalized eigenvalues of the matrix~$\M{R}$ of rectangular plate (Example~\hyperref[tab:Examples]{R2}). Since the matrix~$\Smat$ has only $510$~rows, the number of modes is limited. The number of well-determined modes is delimited by horizontal dashed lines.}
		\label{fig:Rectangle_Rdecomp}
	\end{center}
\end{figure}
}
{}

\ifCLASSOPTIONcaptionsoff
  \newpage
\fi

\bibliographystyle{IEEEtran}
\bibliography{references,total,bibadd}

\begin{thebibliography}{10}
\providecommand{\url}[1]{#1}
\csname url@samestyle\endcsname
\providecommand{\newblock}{\relax}
\providecommand{\bibinfo}[2]{#2}
\providecommand{\BIBentrySTDinterwordspacing}{\spaceskip=0pt\relax}
\providecommand{\BIBentryALTinterwordstretchfactor}{4}
\providecommand{\BIBentryALTinterwordspacing}{\spaceskip=\fontdimen2\font plus
\BIBentryALTinterwordstretchfactor\fontdimen3\font minus
  \fontdimen4\font\relax}
\providecommand{\BIBforeignlanguage}[2]{{%
\expandafter\ifx\csname l@#1\endcsname\relax
\typeout{** WARNING: IEEEtran.bst: No hyphenation pattern has been}%
\typeout{** loaded for the language `#1'. Using the pattern for}%
\typeout{** the default language instead.}%
\else
\language=\csname l@#1\endcsname
\fi
#2}}
\providecommand{\BIBdecl}{\relax}
\BIBdecl

\bibitem{Harrington_FieldComputationByMoM}
R.~F. Harrington, \emph{Field Computation by Moment Methods}.\hskip 1em plus
  0.5em minus 0.4em\relax Wiley -- IEEE Press, 1993.

\bibitem{Sadiku_NumericalTechniques}
M.~N.~O. Sadiku, \emph{Numerical Techniques in Electromagnetics with Matlab},
  3rd~ed.\hskip 1em plus 0.5em minus 0.4em\relax CRC Press, 2009.

\bibitem{GolubVanLoan_MatrixComputations}
G.~H. Golub and C.~F. Van~Loan, \emph{Matrix Computations}.\hskip 1em plus
  0.5em minus 0.4em\relax Johns Hopkins University Press, 2012.

\bibitem{Harrington+Mautz1971}
R.~F. Harrington and J.~R. Mautz, ``Theory of characteristic modes for
  conducting bodies,'' \emph{IEEE Trans. Antennas Propag.}, vol.~19, no.~5, pp.
  622--628, 1971.

\bibitem{YangAdams_SystematicShapeOptimizationOfSymmetricMIMOAntennasUsingCM}
B.~Yang and J.~J. Adams, ``Systematic shape optimization of symmetric mimo
  antennas using characteristic modes,'' \emph{IEEE Trans. Antennas Propag.},
  vol.~64, no.~7, pp. 2668--2678, July 2016.

\bibitem{CapekHazdraEichler_AMethodForTheEvaluationOfRadiationQBasedOnModalApproach}
M.~Capek, P.~Hazdra, and J.~Eichler, ``A method for the evaluation of radiation
  {Q} based on modal approach,'' \emph{IEEE Trans. Antennas Propag.}, vol.~60,
  no.~10, pp. 4556--4567, Oct. 2012.

\bibitem{CapekJelinek_OptimalCompositionOfModalCurrentsQ}
M.~Capek and L.~Jelinek, ``Optimal composition of modal currents for minimal
  quality factor {Q},'' \emph{IEEE Trans. Antennas Propag.}, vol.~64, no.~12,
  pp. 5230--5242, 2016.

\bibitem{GustafssonTayliEhrenborgEtAl_AntennaCurrentOptimizationUsingMatlabAndCVX}
\BIBentryALTinterwordspacing
M.~Gustafsson, D.~Tayli, C.~Ehrenborg, M.~Cismasu, and S.~Norbedo, ``Antenna
  current optimization using {MATLAB} and {CVX},'' \emph{{FERMAT}}, vol.~15,
  no.~5, pp. 1--29, May--June 2016. [Online]. Available:
  \url{http://www.e-fermat.org/articles/gustafsson-art-2016-vol15-may-jun-005/}
\BIBentrySTDinterwordspacing

\bibitem{VogelEtAl_CManalysis_PuttingPhysicsBackIntoSimulation}
M.~Vogel, G.~Gampala, D.~Ludick, U.~Jakobus, and C.~Reddy, ``Characteristic
  mode analysis: Putting physics back into simulation,'' \emph{IEEE Antennas
  Propag. Mag.}, vol.~57, no.~2, pp. 307--317, April 2015.

\bibitem{Chen_2014_UAV_TCM}
Y.~Chen and C.-F. Wang, ``Electrically small {UAV} antenna design using
  characteristic modes,'' \emph{IEEE Trans. Antennas Propag.}, vol.~62, no.~2,
  pp. 535--545, Feb. 2014.

\bibitem{HarringtonMautz_ComputationOfCharacteristicModesForConductingBodies}
R.~F. Harrington and J.~R. Mautz, ``Computation of characteristic modes for
  conducting bodies,'' \emph{IEEE Trans. Antennas Propag.}, vol.~19, no.~5, pp.
  629--639, Sept. 1971.

\bibitem{CapekEtAl_ValidatingCMsolvers}
M.~Capek, V.~Losenicky, L.~Jelinek, and M.~Gustafsson, ``Validating the
  characteristic modes solvers,'' \emph{IEEE Trans. Antennas Propag.}, vol.~65,
  no.~8, pp. 4134--4145, 2017.

\bibitem{Kristensson2016}
G.~Kristensson, \emph{Scattering of Electromagnetic Waves by Obstacles}.\hskip
  1em plus 0.5em minus 0.4em\relax Edison, NJ: SciTech Publishing, an imprint
  of the IET, 2016.

\bibitem{Gustafsson_OptimalAntennaCurrentsForQsuperdirectivityAndRP}
M.~Gustafsson and S.~Nordebo, ``Optimal antenna currents for {Q},
  superdirectivity, and radiation patterns using convex optimization,''
  \emph{IEEE Trans. Antennas Propag.}, vol.~61, no.~3, pp. 1109--1118, 2013.

\bibitem{Harrington_TimeHarmonicElmagField}
R.~F. Harrington, \emph{Time-Harmonic Electromagnetic Fields}, 2nd~ed.\hskip
  1em plus 0.5em minus 0.4em\relax Wiley -- IEEE Press, 2001.

\bibitem{ChewTongHu_IntegralEquationMethodsForElectromagneticAndElasticWaves}
W.~C. Chew, M.~S. Tong, and B.~Hu, \emph{Integral Equation Methods for
  Electromagnetic and Elastic Waves}.\hskip 1em plus 0.5em minus 0.4em\relax
  Morgan \& Claypool, 2009.

\bibitem{Gibson2014}
W.~C. Gibson, \emph{The {M}ethod of {M}oments in {E}lectromagnetics}.\hskip 1em
  plus 0.5em minus 0.4em\relax CRC press, 2014.

\bibitem{Morse+Feshbach1953b}
P.~M. Morse and H.~Feshbach, \emph{Methods of Theoretical Physics}.\hskip 1em
  plus 0.5em minus 0.4em\relax New York, NY: McGraw-Hill, 1953, vol.~2.

\bibitem{Waterman1971}
P.~C. Waterman, ``Symmetry, unitarity, and geometry in electromagnetic
  scattering,'' \emph{Phys. Rev. D}, vol.~3, no.~4, pp. 825--839, 1971.

\bibitem{Hansen1988}
J.~E. Hansen, Ed., \emph{Spherical Near-Field Antenna Measurements}, ser. {IEE}
  electromagnetic waves series.\hskip 1em plus 0.5em minus 0.4em\relax
  Stevenage, UK: Peter Peregrinus Ltd., 1988, no.~26.

\bibitem{Gustafsson+Nordebo2006c}
M.~Gustafsson and S.~Nordebo, ``Characterization of {MIMO} antennas using
  spherical vector waves,'' \emph{IEEE Trans. Antennas Propag.}, vol.~54,
  no.~9, pp. 2679--2682, 2006.

\bibitem{Waterman1965}
P.~Waterman, ``Matrix formulation of electromagnetic scattering,'' \emph{Proc.
  IEEE}, vol.~53, no.~8, pp. 805--812, Aug. 1965.

\bibitem{Liu+etal2012}
Y.~G. Liu, W.~C. Chew, L.~Jiang, and Z.~Qian, ``A memory saving fast {A-EFIE}
  solver for modeling low-frequency large-scale problems,'' \emph{Applied
  Numerical Mathematics}, vol.~62, no.~6, pp. 682--698, 2012.

\bibitem{Zuras2008}
D.~Zuras, M.~Cowlishaw, A.~Aiken, M.~Applegate, D.~Bailey, S.~Bass,
  D.~Bhandarkar, M.~Bhat, D.~Bindel, S.~Boldo \emph{et~al.}, ``{IEEE} standard
  for floating-point arithmetic,'' \emph{IEEE Std 754-2008}, pp. 1--70, 2008.

\bibitem{Burden2015}
R.~Burden, J.~Faires, and A.~Burden, \emph{Numerical Analysis}.\hskip 1em plus
  0.5em minus 0.4em\relax Cengage Learning, 2015.

\bibitem{Song+Chew2001a}
J.~Song and W.~C. Chew, ``Error analysis for the truncation of multipole
  expansion of vector green's functions [em scattering],'' \emph{IEEE Microwave
  and Wireless Components Letters}, vol.~11, no.~7, pp. 311--313, 7 2001.

\bibitem{mpmath2013}
\BIBentryALTinterwordspacing
F.~Johansson \emph{et~al.} (2013, December) mpmath: a {P}ython library for
  arbitrary-precision floating-point arithmetic (version 0.18). [Online].
  Available: \url{http://mpmath.org/}
\BIBentrySTDinterwordspacing

\bibitem{IntelMKL2017}
\BIBentryALTinterwordspacing
Intel. (2017) Intel {M}ath {K}ernel {L}ibrary 2017 update 3. [Online].
  Available: \url{https://software.intel.com/en-us/mkl}
\BIBentrySTDinterwordspacing

\bibitem{RaoWiltonGlisson_ElectromagneticScatteringBySurfacesOfArbitraryShape}
S.~M. Rao, D.~R. Wilton, and A.~W. Glisson, ``Electromagnetic scattering by
  surfaces of arbitrary shape,'' \emph{IEEE Trans. Antennas Propag.}, vol.~30,
  no.~3, pp. 409--418, May 1982.

\bibitem{atom}
\BIBentryALTinterwordspacing
(2017) {A}ntenna {T}oolbox for {MATLAB} ({AToM}). Czech Technical University in
  Prague. [Online]. Available: \url{www.antennatoolbox.com}
\BIBentrySTDinterwordspacing

\bibitem{IDA}
D.~Tayli. (2017) {IDA} ({I}ntegrated {D}evelopment toolset for {A}ntennas).
  Lund University.

\bibitem{feko}
\BIBentryALTinterwordspacing
Altair. (2016) {FEKO}. Altair. [Online]. Available: \url{www.feko.info}
\BIBentrySTDinterwordspacing

\bibitem{advanpix}
\BIBentryALTinterwordspacing
{A}dvanpix. (2016) {M}ultiprecision {C}omputing {T}oolbox for {MATLAB}.
  [Online]. Available: \url{http://www.advanpix.com/}
\BIBentrySTDinterwordspacing

\bibitem{Dunavant_HighDegreeEfficientGQR}
D.~A. Dunavant, ``High degree efficient symmetrical gaussian quadrature rules
  for the triangl,'' \emph{International Journal for Numerical Methods in
  Engineering}, vol.~21, pp. 1129--1148, 1985.

\bibitem{SchabBernhard_RadiationAndEnergyStorageCurrentModesOnConductingStructures}
K.~R. Schab and J.~T. Bernhard, ``Radiation and energy storage current modes on
  conducting structures,'' \emph{IEEE Trans. Antennas Propag.}, vol.~63,
  no.~12, pp. 5601--5611, Dec. 2015.

\bibitem{Wilkinson_AlgebraicEigenvalueProblem}
J.~H. Wilkinson, \emph{The Algebraic Eigenvalue Problem}.\hskip 1em plus 0.5em
  minus 0.4em\relax Oxford University Press, 1988.

\bibitem{AngiulliVenneri_SDTforCM}
G.~Angiulli and F.~Venneri, ``Use of the simultaneous diagonalization technique
  in the {$Ax = \lambda Bx$} eigenproblem applied to the computation of the
  characteristic modes,'' \emph{ACES Journal}, vol.~17, no.~3, pp. 232--238,
  Nov. 2002.

\bibitem{Capek+etal2017b}
M.~Capek, M.~Gustafsson, and K.~Schab, ``Minimization of antenna quality
  factor,'' \emph{IEEE Trans. Antennas Propag.}, vol.~65, no.~8, pp.
  4115–--4123, 2017.

\bibitem{JelinekCapek_OptimalCurrentsOnArbitrarilyShapedSurfaces}
L.~Jelinek and M.~Capek, ``Optimal currents on arbitrarily shaped surfaces,''
  \emph{IEEE Trans. Antennas Propag.}, vol.~65, no.~1, pp. 329--341, Jan. 2017.

\bibitem{nastran}
\BIBentryALTinterwordspacing
(2017) {MSC NASTRAN}. [Online]. Available:
  \url{http://www.mscsoftware.com/support/}
\BIBentrySTDinterwordspacing

\bibitem{EibertHansen_OnTheCalculationOfPotenticalIntegralsForLinearSourceDistributionsOnTriangularDomains}
T.~F. Eibert and V.~Hansen, ``On the calculation of potential integrals for
  linear source distributions on triangular domains,'' \emph{IEEE Trans.
  Antennas Propag.}, vol.~43, no.~12, pp. 1499--1502, Dec. 1995.

\bibitem{DEMCEM2010}
\BIBentryALTinterwordspacing
A.~G. Polimeridis. (2010) Direct evaluation method in computational
  electromagnetics ({DEMCEM}). [Online]. Available:
  \url{https://github.com/thanospol/DEMCEM}
\BIBentrySTDinterwordspacing

\bibitem{Polimeridis+Yioultsis2008}
A.~G. Polimeridis and T.~V. Yioultsis, ``On the direct evaluation of weakly
  singular integrals in galerkin mixed potential integral equation
  formulations,'' \emph{IEEE Trans. Antennas Propag.}, vol.~56, no.~9, pp.
  3011--3019, 2008.

\bibitem{Polimeridis+Wilton2010}
A.~G. Polimeridis and J.~R. Mosig, ``Complete semi-analytical treatment of
  weakly singular integrals on planar triangles via the direct evaluation
  method,'' \emph{International journal for numerical methods in engineering},
  vol.~83, no.~12, pp. 1625--1650, 2010.

\bibitem{Polimeridis+Mosig2011}
------, ``On the direct evaluation of surface integral equation impedance
  matrix elements involving point singularities,'' \emph{IEEE Trans. Antennas
  Propag.}, vol.~10, pp. 599--602, 2011.

\bibitem{Dagum1998}
L.~Dagum and R.~Menon, ``Open{MP}: an industry standard api for shared-memory
  programming,'' \emph{Computational Science \& Engineering, IEEE}, vol.~5,
  no.~1, pp. 46--55, 1998.

\bibitem{Olver+etal2010}
F.~W.~J. Olver, D.~W. Lozier, R.~F. Boisvert, and C.~W. Clark, \emph{{NIST}
  {H}andbook of mathematical functions}.\hskip 1em plus 0.5em minus 0.4em\relax
  New York: Cambridge University Press, 2010.

\bibitem{Jeffrey_MathHandbook}
A.~Jeffrey and H.-H. Dai, \emph{Handbook of Mathematical Formulas and
  Integrals}, 4th~ed.\hskip 1em plus 0.5em minus 0.4em\relax Academic Press,
  2008.

\end{thebibliography}

\begin{IEEEbiography}[{\includegraphics[width=1in,height=1.25in,clip,keepaspectratio]{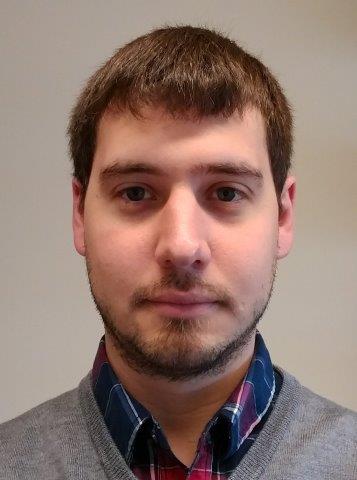}}]{Doruk Tayli}(S'13) 
received his B.Sc. degree in Electronics Engineering from Istanbul Technical University and his M.Sc. in degree in Communications Systems from Lund University, in 2010 and 2013, respectively. He is currently a Ph.D. student at Electromagnetic Theory Group, Department of Electrical and Information Technology at Lund University. 
	
His research interests are Physical Bounds, Small Antennas and Computational Electromagnetics.
\end{IEEEbiography}

\begin{IEEEbiography}[{\includegraphics[width=1in,height=1.25in,clip,keepaspectratio]{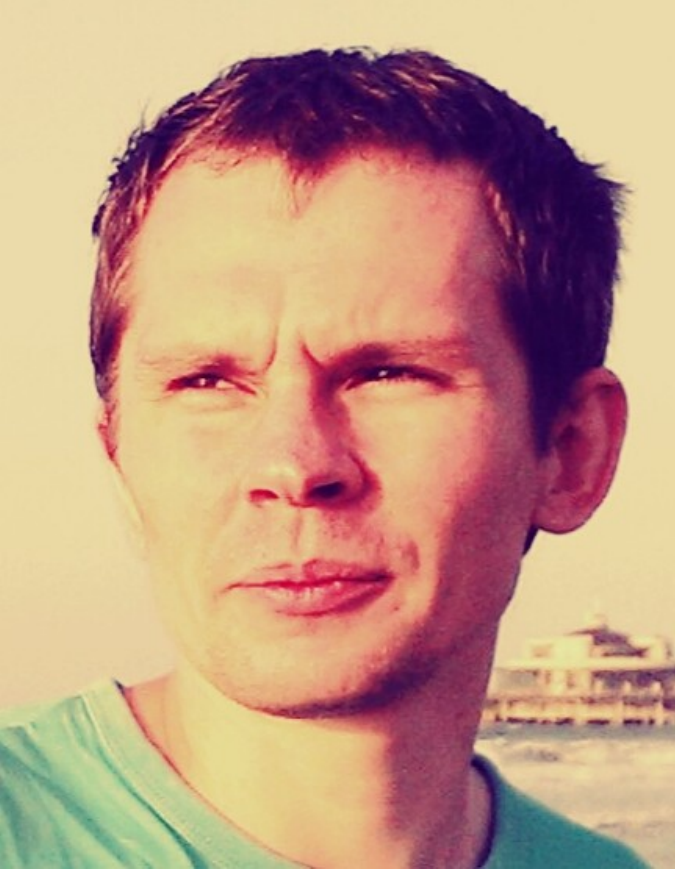}}]{Miloslav Capek}(SM'17)
received his Ph.D. degree from the Czech Technical University in Prague, Czech Republic, in 2014. In 2017 he was appointed Associate Professor at the Department of Electromagnetic Field at the same university.

He leads the development of the AToM (Antenna Toolbox for Matlab) package. His research interests are in the area of electromagnetic theory, electrically small antennas, numerical techniques, fractal geometry and optimization. He authored or co-authored over 70 journal and conference papers.

Dr. Capek is member of Radioengineering Society, regional delegate of EurAAP, and Associate Editor of Radioengineering.
\end{IEEEbiography}

\begin{IEEEbiography}[{\includegraphics[width=1in,height=1.25in,clip,keepaspectratio]{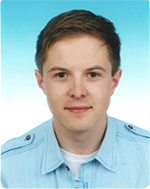}}]{Vit Losenicky}
received the M.Sc. degree in electrical engineering from the Czech Technical University in Prague, Czech Republic, in 2016. He is now working towards his Ph.D. degree in the area of electrically small antennas.
\end{IEEEbiography}

\begin{IEEEbiography}[{\includegraphics[width=1in,height=1.25in,clip,keepaspectratio]{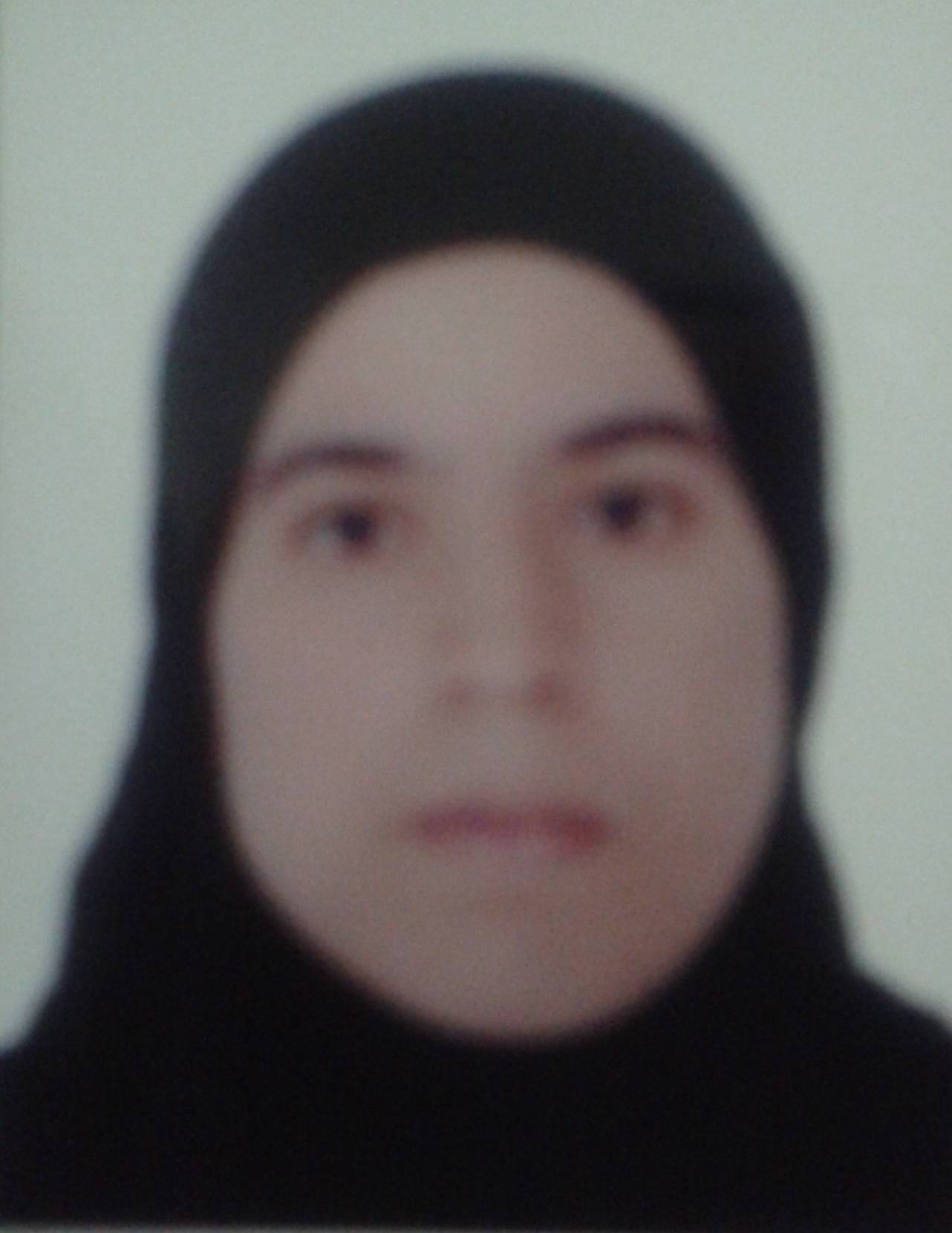}}]{Akrou Lamyae}
received the Dipl.-Ing. degree in networks and telecommunications
 from National School of Applied Sciences of Tetouan in 2012. Since 2014 she is working towards her Ph.D. degree in Electrical and Computer Engineering at the University of Coimbra.

\end{IEEEbiography}

\begin{IEEEbiography}[{\includegraphics[width=1in,height=1.25in,clip,keepaspectratio]{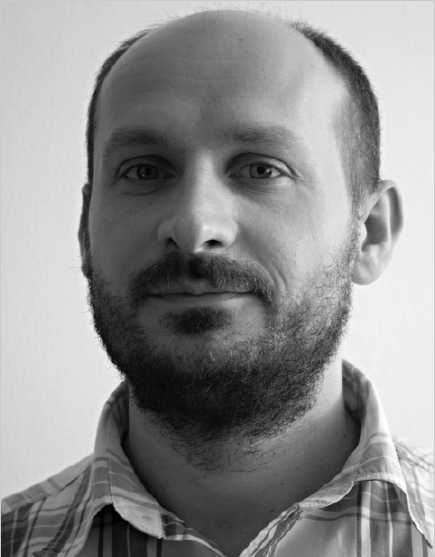}}]{Lukas Jelinek}
received his Ph.D. degree from the Czech Technical University in Prague, Czech Republic, in 2006. In 2015 he was appointed Associate Professor at the Department of Electromagnetic Field at the same university.

His research interests include wave propagation in complex media, general field theory, numerical techniques and optimization.
\end{IEEEbiography}

\begin{IEEEbiography}[{\includegraphics[width=1in,height=1.25in,clip,keepaspectratio]{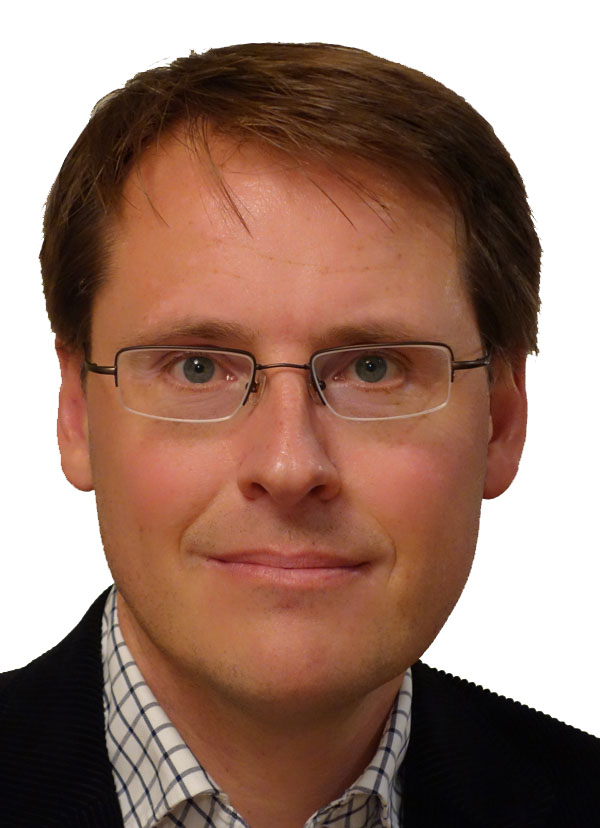}}]{Mats Gustafsson }(SM'17)
received the M.Sc. degree in Engineering Physics 1994, the Ph.D. degree in Electromagnetic Theory 2000, was appointed Docent 2005, and Professor of Electromagnetic Theory 2011, all from Lund University, Sweden. 

He co-founded the company Phase holographic imaging AB in 2004. His research interests are in scattering and antenna theory and inverse scattering and imaging. He has written over 90 peer reviewed journal papers and over 100 conference papers. Prof. Gustafsson received the IEEE Schelkunoff Transactions Prize Paper Award 2010 and Best Paper Awards at EuCAP 2007 and 2013. He served as an IEEE AP-S Distinguished Lecturer for 2013-15.
\end{IEEEbiography}

\end{document}